\documentclass{aa}
\input psfig.tex
\usepackage{graphicx}
\usepackage{natbib}

\usepackage{array}
\usepackage{graphics}
\usepackage{latexsym}
\usepackage{amssymb}
\usepackage{amsmath}
\usepackage{fancyhdr}
\usepackage{morefloats}
\usepackage{bm}
\bibpunct{(}{)}{;}{a}{}{,}

\begin{document}

\title{Dissipative phenomena in extended-bodies interactions I: Methods}
\subtitle{ Dwarf galaxies of the Local Group and their synthetic CMDs}

\author{S.\ Pasetto\inst{1,2},
			  G. Bertelli\inst{3},
				E.K.\ Grebel\inst{1},
				C. Chiosi\inst{4},
				Y. Fujita\inst{5}
           }

\offprints{sm2@mssl.ucl.ac.uk}

\institute{Astronomisches Rechen-Institut, Zentrum f\"ur Astronomie der Universit\"at Heidelberg, Heidelberg, Germany
\and
University College London, Department of Space \& Climate Physics, Mullard Space Science Laboratory, Holmbury St. Mary, Dorking Surrey RH5 6NT, United Kingdom
\and
INAF - Padova Astronomical Observatory, Padova, Italy
\and
Astronomy Department, Padova University, Padova, Italy
\and
Department of Earth and Space Science, Graduate School of Science, Osaka University,
Toyonaka, Osaka, Japan
}
\date{Received: July 2011;  Accepted on A\&A}

\titlerunning{Dissipative phenomena in LG dwarf galaxies}
\authorrunning{S.\ Pasetto et al.}

\abstract {}
{Dissipative phenomena occurring during the orbital evolution of a dwarf satellite galaxy around a host galaxy may leave signatures in the star formation activity and signatures in the colour magnitude diagram of the galaxy stellar content. Our goal is to reach a simple and qualitative description of the these complicated phenomena.}
{We develop an analytical and numerical technique able to study ram pressure, Kelvin-Helmholtz instability, Rayleigh-Taylor and tidal forces acting on the star formation processes in molecular clouds. We consider it together with synthetic colour magnitude diagrams techniques.}
{We developed a method to investigate the connections existing between gas consumption processes and star formation processes in the context of the two extended-body interaction with special attention to the dwarf galaxies dynamical regime.} {}


\maketitle

\section{Introduction}\label{Introduction}
We investigate the role of dissipative phenomena in the star formation history of a dwarf satellite galaxy orbiting around a primary galaxy. Our purpose is to quantify in a simple way the connections between the gas consumption processes (e.g., ram pressure, Rayleigh-Taylor and Kelvin-Helmholtz, tidal forces) and the star formation processes in the context of the interaction between two extended-bodies with special attention to the  dynamical regime of dwarf galaxies.

Dwarf galaxies offer a beautiful example of interactions between a major galaxy and its satellites \citep[e.g.,][]{1993ppc..book.....P, 1980lssu.book.....P}. While orbiting these satellites are affected by environmental effects that shape their evolution and may affect their star formation history \citep[e.g.,][]{2003AJ....125.1926G}. Their star formation history can be inferred from their colour magnitude diagrams (CMD), see, e.g. the recent analysis by \citet{2008ApJ...686.1030O} of Hubble Space Telescope data for the galaxies of the Local Group (LG). Generally, environmental effects are important in groups and clusters of galaxies and are believed to contribute to the origin of the density-morphology relation \citep[e.g.,][]{1974ApJ...194....1O, 1980ApJ...236..351D, 1997ApJ...490..577D} or to the Butcher-Oemler effect \citep{1978ApJ...219...18B, 1998ApJ...497..188C,  2003PASJ...55..757G}.  The mechanism of tidal compression  exerted by the gravitational potential of a galaxy cluster  is known to increase the velocity dispersion of the molecular clouds falling in disk galaxies \citep[e.g.,][]{1993ApJ...408...57V, 1990ApJ...350...89B} thus inducing star formation episodes \citep[e.g.,][]{1996ApJ...459...82H, 1991MNRAS.248P...8E}. Tidal compression is also responsible for the  so-called galaxy-harassment phenomenon in the case of  dwarf  galaxies \citep{1996Natur.379..613M, 1998ApJ...495..139M, 1999MNRAS.304..465M} or  late-type systems \citep[e.g.,][]{2003ApJ...582..141G}. In this context, three-dimensional numerical simulations \citep[e.g.,][]{2000Sci...288.1617Q} show that ram pressure can be effective in removing up to $100\% $  of the atomic hydrogen from luminous galaxies within $10^8$ yr.

With respect to the LG dwarf galaxies, the role of the external mechanisms of gas removal has been investigated by means of  different approaches and  with conflicting results \citep[for a review see][]{2003AJ....125.1926G} and in relation with the MW we cannot neglect the dissipative role of tidal forces by the deep MW potential and its hot coronal plasma.
A large amount ($ \sim 5 \times 10^{10} M_ \odot  $) of hot gas is thought to exist in an extended ($ \sim 200$kpc) hot diffuse halo  \citep[see e.g.,][]{2000A&A...357..120B, 2003ApJS..146..165S, 2003ApJ...597..948P, 2004MNRAS.355..694M, 2006ApJ...639..590F, 2006MNRAS.370.1612K, 2006ApJ...644L...1S}. Ram pressure can both enhance star formation by compression of the interstellar medium (ISM) \citep{1999ApJ...516..619F} and reduce it by stripping the gas from the galaxy \citep{1972ApJ...176....1G}. Also, tidal forces can trigger episodes of star formation. These are the major external competitive effects that we want to investigate in the context of 'dwarf galaxy-massive galaxy' interaction. In addition to this, there are also several internal mechanisms that govern the gas consumption, such as gas expulsion by star formation processes \citep[e.g.,][]{ 1997ApJ...480..297B, 2000MNRAS.313..291F, 1998LNP...506..559M} giant HII regions  \citep[e.g.,][]{ 1988MNRAS.233..553D, 2000ApJ...540..814E}. The role of the  hot gaseous halo surrounding the MW  has often been  studied in relation with the Magellanic Stream, a long HI filament leading and trailing the Magellanic Clouds   \citep[see e.g.,][]{2000ApJ...529L..81M,2010ApJ...723.1618N}, where the authors constrained the density of the coronal gas to be $n < 10^{ - 5} {\rm{cm}}^{ - 3} $ at the distance of 50 kpc in order to let the Stream clouds survive up to 2.5 Gyr or at least a minimum of 0.5 Gyr.  \citet{2002ApJ...576..773S} suggested a limit of $n \cong 3 \cdot 10^{ - 4} {\rm{cm}}^{ - 3} $ at the same distance, which is essentially in agreement with the more recent measurements  by \citet{2009ApJ...698.1485H}. Previous determinations by \citet{1998A&A...339..745K} based  on the ROSAT X-ray observations agreed on the same limit for the MW halo, whereas  models  predicting dwarf galaxies to lose most of their gas
\citep[e.g.,][]{2001AJ....121.2572G, 1974Natur.252..111E} need a gas density one order of magnitude higher.
On the other hand, \citet{2009ApJ...696..385G} argue that ram pressure is one of the primary causes for the gas deficiency of dwarf satellites within 270 kpc around the MW and M31 \citep[see also][]{1994ApJ...428..617V, 2003AJ....125.1926G}. Finally, we assume that the present-day density of the intergalactic medium beyond the LG halos is too low for ram pressure to be effective \citep[e.g.,][]{ 2001ApJ...555L..95Q}.

The subject of our analysis will be an arbitrary idealized dwarf galaxy orbiting around the Milky Way (MW) galaxy. The differential equations involved in our
study are generally coupled in a non-linear way thus not allowing us to consider the global evolution of the processes as a linear superposition of the single effects considered alone and then summed up. Nevertheless, for the sake of clarity, it is customary to subdivide the effects involved in the evolution of a dwarf galaxy in two categories: 'internal' and 'external'; and then to consider their superposition. We will follow this tradition also in this paper. Among the internal effects, stellar evolution processes are the most important ones because they ultimately govern the location of the stars in a CMD.
The evolution of the stars and their feedback is mainly related to their mass and chemical composition. In our approach, we will carefully consider stellar feedback on the dwarf galaxy stellar populations in dependence of the mass of the stars (see Section \ref{Molecular clouds evolution and non gravitational heating}).
Finally, in our approximation we want to consider several processes of external nature as well, mainly tidal compression, ram pressure, Rayleigh-Taylor and Kelvin-Helmholtz instabilities.
The reasons for focusing on these effects are the following.
We are mostly interested in the effects of the orbits on the generation of observable CMD. To approach this problem we will work out a spherical approximation for the shape of the idealized dwarf galaxy considered as a whole. On the one hand, this strong approximation will make us lose track of the orbital evolution of the stars inside the galaxy itself, i.e. we will not solve the equation of motion (EoM) for individual stars inside the dwarf galaxy (see Section \ref{Tidal approximation for a pressure equation in non-inertial reference systems}), but this approach will allow us to neglect the effects of tidal stripping and tidal stirring \citep[e.g.,][]{2001ApJ...559..754M, 2003A&A...405..931P, 2006MNRAS.367..387R, 2009MNRAS.397.2015K} giving more agility to our code (see the code performance in Appendix A) as compared with standard N-body techniques \citep[see, e.g.,][for a recent review]{2010AdAst2010E..25M}.
Finally, in order to simplify further the orbit integration, we will limit ourself to non-penetrating encounters with the MW, i.e., we will limit our analysis to the  dwarf galaxy orbit pericenter, $\bm{r}_p$, greater than $\left\| {{\bm{r}}_p } \right\| > 20$ kpc, thus neglecting partially the problem of the indetermination of the parameters of the MW potential, the role of the dynamical friction and role of tidal shocks due to the passages trough the MW disk (which are not the target of the present paper).
Being mainly focused on the LG dwarf galaxy orbits, we will not consider galaxy harassment because it is much less efficient in low density groups such as our LG \citep[e.g.,][]{1996Natur.379..613M}.

The paper is organized as follow: In Section \ref{A formulation for gas consumption and star formation for dwarf galaxies} we present our model for gas the consumption and star formation in dwarf galaxies taking into account both internal and environmental effects that would affect the star formation history while the dwarf galaxy is orbiting about a hosting galaxy (the MW in this paper). In
Section \ref{Synthetic colour magnitude diagrams} we shortly summarize the key lines of the population synthesis technique to generate colour-magnitude diagrams out of which the history of star formation is inferred. In Section \ref{An application to the local group} the technique is applied to a model of a dwarf galaxy in which the star formation efficiency is altered by the environmental effects during the orbital motion and from which a synthetic CMD is generated and shortly analysed. A brief discussion of the results and some general conclusions are presented in Section \ref{Conclusions}.

Finally, we made extensive use of the Appendix to present technical test on the code or to explain the analytical aspects of the work that can be postponed at a first reading of the paper.

\section{A model for gas consumption and star formation in dwarf galaxies}\label{A formulation for gas consumption and star formation for dwarf galaxies}
\subsection{Molecular clouds evolution and non gravitational heating}\label{Molecular clouds evolution and non gravitational heating}

To link the orbital motion of a satellite about its host galaxy  to the environmental effects and the star formation efficiency in the satellite itself, we must set up  a fast algorithm allowing us to explore a large space of initial conditions for the orbital motions. This immediately rules out a multi-phase fully hydrodynamic treatment of the gas  \citep[e.g.,][]{2000Sci...288.1617Q, 2007ApJ...671.1434T} because it would be too time consuming in particular if a large volume of initial conditions (i.c.) of the orbital motions has to be explored.  Therefore, we prefer to follow the  method proposed by \citet{1998ApJ...509..587F} and \citet{1999ApJ...516..619F} from whom we take the model of star formation to be integrated by extensive calculations of orbital motions and synthetic CMDs of the stellar content of the dwarf galaxy.

In brief, stars are known to form in gas clouds (either atomic or molecular) whose mass distribution  is uncertain but believed to span at least a factor $ \sim 10^6 $ in mass \citep{1997ApJ...480..235E}. We group  the molecular clouds according to their initial mass $M_i |M_i  \in \left[ {M_{\min },M_{\max } } \right]$  and  assume $M_{\min }  \cong 10^2 M_ \odot  $  and $M_{\max }  \cong 10^8 M_ \odot  $. Since this mass interval is very large, we split it logarithmically and resolve the mass intervals with a ratio $\alpha $ between any two subsequent intervals, i.e. we solve the recurrence relation $\log _{10} \frac{{M_{i + 1} }}{{M_i }} = \alpha $,  with the boundary condition $M_0  = M_{\min } $ for i=0, as $M_i  = 10^{\alpha i}  M_{\min } $ ($\alpha$ is a rational number). The numerical convergence is confirmed for $\alpha = \frac{1}{{100}}$ (see Appendix A for details).

We then define for simplicity a ``mass class'' within the generic interval $\hat M_i  \equiv M_{i + 1}  - M_i $
or its fraction over a total mass $M_{{\rm{tot}}} $ as $\hat f_i  \equiv {\raise0.7ex\hbox{${\hat M_i }$} \!\mathord{\left/
{\vphantom {{\hat M_i } {M_{\text{tot}} }}}\right.\kern-\nulldelimiterspace}
\!\lower0.7ex\hbox{${M_{\text{tot}} }$}}$. We assume that the evolution of each mass class $\hat M_i $ representative of each mass interval is governed by  the fraction of gas that is ejected from the stars, say their gas ejection rate $R_{{\rm{star}}} $ and by the amount of recycled molecular gas with recycling rate $R_{{\rm{mol}}} $.

To determine $R_{{\rm{star}}} $ we remember that short lived-lived and long-lived stars contribute to  $R_{{\rm{star}}} $ in a different fashion. If  $r\left( m \right)$ is the amount of mass returned by a star of mass $m$,     $\mu \left( m \right) = \frac{{r\left( m \right)}}{m}$ is  the fractionary return mass. Then the gas ejection rate due to stars with short lifetime, $R_{{\rm{star}}}^{{\rm{[S]}}} $, will depend on the mass distribution of the stars, i.e. the time independent initial mass function $\iota  = \iota \left( m \right)$ expressed as a mass fraction, and the star formation rate $\psi $ at the moment in which the star is born $\psi \left( {t - \tilde t\left( m \right)} \right)$ with $\tilde t\left( m \right)$
 being the lifetime of a star of mass $m$. Supposing that the typical age of a galaxy is, for instance, $t_G = 12$ Gyr and looking for mass of stars whose lifetime is small compared to the age of the galaxy (say 1/10 $t_G$),  $R_{{\rm{star}}}^{{\rm{[S]}}}$ is given by

\begin{equation}\label{eq2}
R_{{\rm{star}}}^{{\rm{[S]}}} \left( t \right) = \int_{m_{{\rm{low}}} }^{m_{{\rm{up}}} } {\psi \left( {t - \tilde t\left( m \right)} \right)\mu \left( m \right)\iota \left( m \right)dm},
\end{equation}
where
 $m_{{\rm{low}}}  = 2.3M_ \odot  $ is the star with $\tilde t\left( m \right) \simeq 0.1 t_G$ and $m_{up}$ is the most massive star we want to consider $m_{{\rm{up}}}  = 100M_ \odot  $. All the less massive stars contribute to the long-lived part of $R_{{\rm{star}}} $ indicated as  $R_{{\rm{star}}}^{{\rm{[L]}}} $. The contribution of  these stars can be written as  $R_{{\rm{star}}}^{{\rm{[L]}}}  = \psi_0  - R_{{\rm{star}}}^{{\rm{[S]}}} \left( {0 } \right)$. $\psi_0 $ is an artificially fixed star formation rate for the galaxy we are considering and we consider the lifetime corresponding to the transition mass from long-lived to short lived stars $t_{tr}  = 0.1 t_G$ Gyr. Recollecting our previous terms, we get
\begin{align}
\begin{array}{l}
 R_{{\rm{star}}}  = R_{{\rm{star}}}^{{\rm{[S]}}}  + R_{{\rm{star}}}^{{\rm{[L]}}}  \\
  = \theta \left( t \right)\left( {R_{{\rm{star}}}^{{\rm{[S]}}} \left( t \right) - R_{{\rm{star}}}^{{\rm{[S]}}} \left( 0 \right)} \right) + \psi_0  \\
  = \int_{m_{{\rm{low}}} }^{m_{{\rm{up}}} } {\left[ {\psi \left( {t - \tilde t} \right) - \psi \left( { - \tilde t} \right)} \right] \mu(m) \iota(m) dm}, \\
 \end{array}\label{eq3}
\end{align}
where $\theta \left( x \right)$ is a step function ($\theta  = 0$ for $x \le 0$, $\theta  = 1$ otherwise), and  $R_{{\rm{star}}}^{{\rm{[L]}}}$ is indeed not time dependent for $t>0$.

Furthermore, we  define the molecular recycling rate, $R_{{\rm{mol}}} $, as the complement of the star formation efficiency, $1 - \varepsilon $ where $\varepsilon $
is a general function  of the molecular cloud mass $M$ and the pressure $P$,  i.e. $\varepsilon  = \varepsilon \left( {M,P} \right)$. The star formation efficiency in turn can be reduced by the presence of an external gas consumption mechanism, e.g., HI phase, ram pressure, Rayleigh-Taylor instability, tidal compression etc. (see the Section \ref{External gas consumption effect} below  for a detailed discussion). Therefore, for each mass class, $i$,  we  can define
\begin{equation}\label{eq4}
R_{\text{mol}} \left( t \right) = \sum\limits_i^{} {\left( {1 - \varepsilon \left( {M_i ,P} \right)} \right)} \frac{{\hat M_i \left( t \right)}}{{\tau \left( {M_i ,P} \right)}}
\end{equation}
which depends on all the molecular clouds $i$, and where $\tau \left( {M_i ,P}\right)$ in the destruction time of a molecular cloud with mass $M_i$ and pressure $P$ (see below for an extended discussion, and Appendix B for the analytical formulation). In principle, it is now simple to introduce in Eqn. \eqref{eq4} a reducing factor in $R_{{\rm{mol}}} $ in order to lower the fraction $1 - \varepsilon _{{\rm{eff}}} $ when the stronger energy injection by supernovae, photo-ionization of molecular clouds, or the effects of strong stellar winds need to be investigated. \textsl{We have chosen to focus the present work mostly on external effects on the gas consumption or gas removal.}

Finally, the time variation of $\hat M_i $  simply depends on the life-time  $\tau$ of the molecular cloud according to the system of integro-differential equations\footnote{To reduce the degrees of complexity of the equations, we will not consider cosmological effects or the explicit dependence of the equations on the redshift.}:

\begin{equation}\label{eq1}
	\frac{{d\hat M_i }}{{dt}} = \theta \left( {\hat M_i } \right)\left[ {\hat f_i \left( {R_{\text{star}}  + R_{\text{mol}} } \right) - \frac{{\hat M_i }}{\tau }} \right],
\end{equation}
which relates the rate of change of each mass class $\frac{{d\hat M_i }}{{dt}}$ to all the other molecular classes though the gas ejection rate of the stars $R_{{\text{star}}} $, and the recycle rate of molecular clouds $R_{{\text{mol}}} $ (where we again use the step function defined after Eqn. \eqref{eq3} to take into account that the mass is a positive definite quantity).

Once the system of equations \eqref{eq1} is considered together with Eqns. \eqref{eq2}, \eqref{eq3} and \eqref{eq4}, we can estimate the star formation rate (SFR) as
\begin{equation}\label{eq5}
\psi \left( t \right) = \sum\limits_i^{} {\varepsilon \left( {M_i ,P} \right)\frac{{\hat M_i \left( t \right)}}{{\tau \left( {M_i ,P} \right)}}},
\end{equation}
which relates the evolution of the $i^{th}$ molecular cloud with all the other $N-1$ molecular clouds as seen in Eqn \eqref{eq2} and represents the target of our computation.

In the previous system of equations the only functions that needs to be adopted from the literature are the star formation efficiency of a molecular cloud $\varepsilon \left( {M_i ,P} \right)$ and the destruction time $\tau \left( {M_i ,P}\right)$ of a molecular cloud with mass $M_i$ and pressure $P$. Then, once these function are assumed, the system can be integrated once $\psi_0$ is assumed on the basis of the system we want to analyse (e.g., globular clusters, dwarf galaxies, spiral galaxies etc.).

\begin{figure}
\resizebox{\hsize}{!}{\includegraphics{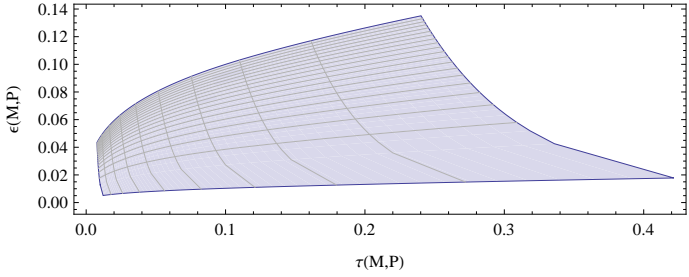}}
\caption{Star formation efficiency vs. lifetime relation. The surface is an interpolation with 20 meshes of constant pressure run horizontally and linearly for $P \in \left[{0.001,1}\right]P_\odot$ from top to bottom. Ten mesh lines runs vertically for constant mass, logarithmically for $\hat M_i  \in \left[ {10^2 ,10^6 } \right]M_ \odot  $ from left to right.}
\label{Complex}
\end{figure}

Star formation most likely responds  to the pressure variation in the local interstellar medium, and in our case it is worth recalling that the pressure $P = P\left( {{\bm{x}},t} \right)$ depends on time and relative position with respect to the host galaxy (with the barycentre centred in the origin $O$ of the inertial reference system $S_0$ with vector radius $x$) and the environment.
 The relation between star formation efficiency, $\varepsilon $, mass $M$ and pressure $P$ of molecular clouds has been studied e.g., in \citet{1997ApJ...480..235E} and \citet{2005ApJ...630..250K} showing how  a pressure increase causes  an increment of $\varepsilon $ and a subsequent decrease of the lifetime of  a cloud $\tau $. These relations are independent of the physical origin of the pressure and can be safely applied to our treatment of the SFH during the orbital motion of a dwarf galaxy. We simply assume these relations from the literature. A plot of their values can be seen in Fig. \ref{Complex} for the range of values adopted in the case of the dwarf orbiting around the MW. We refer the interested reader to further details and their analytical definition in Appendix B.

Finally, we mention that the dynamical response to gas removal by SN explosions in dwarf galaxies has been repeatedly invoked as a mechanism to suppress the star formation in these objects  \citep[e.g.,][]{1999MNRAS.310.1087S, 2000MNRAS.319..168C,
2004MNRAS.349.1101D, 2004ApJ...610...23N, 2005ApJ...631...21K}.
These results are nevertheless in contrast with observational properties as pointed out in \citet{1999ApJ...513..555S}, \citet{2000ApJ...538..477N}, \citet{2000ApJ...528..607N}, \citet{2007MNRAS.375..913P} suggesting that numerical recipes can be improved as, e.g., recently shown by \citet[][]{2010A&A...514A..47P} using prescriptions based on chemo-dynamical studies. These authors made evident that the gas survival in spherical dwarf galaxies is perfectly compatible with the expected SN rate over the whole mass range of interest for dwarf galaxies (see their Table 1). This reasonably stems  from the reduced SN feedback expected in the local environment \citep[][]{1991ApJ...367...96C} different from dwarf to spiral galaxies \citep{1998A&A...337..338B} and it explains in a natural way the co-existence of old stellar populations and a reservoir of neutral gas \citep[e.g.,][]{1990MNRAS.244..168P,2010A&A...514A..47P}.
%%%%%%%%%%%%%%%%%%%%%%Figure 0

\subsection{Galaxy model and plasma distribution}\label{Galaxy model and Plasma distribution}

The most important contributions to the understanding of the plasma corona of the MW come from particle physics \citep[e.g.,][]{2000ApJ...529L..81M} and  cosmology \citep[e.g.,][]{2006ApJ...639..590F}. Nevertheless, the literature is rich in attempts to model the  distributions of the hot gas density (which because of the high degree of ionization refers uniquely to electron density) in different environments
 e.g. in the LG \citep{2001ApJ...559..892R, 2001MNRAS.327L..27B, 1999ApJ...522L..81M} by considering different constraints, in  clusters of galaxies by fitting the X-ray observations with $\beta $
-models \citep[e.g.,][]{1976A&A....49..137C} and by fitting the MW hot coronal gas \citep[e.g.,][]{1994MNRAS.270..209M, 2006ApJ...639..590F,2008SSRv..134...25R,2003ApJS..146..125S,2005ApJ...623L..97T}. In a hierarchical galaxy formation scenario, after the initial collapse, the continuing accretion of gas-rich fragments can produce a diffuse hot gas halo that surrounds the galaxies and fills the dark matter potential \citep{1978MNRAS.183..341W, 1991ApJ...379...52W} not necessarily with the same radial profiles \citep{2006ApJ...639..590F}. This scenario agrees also with the case of clusters of galaxies where the X-ray distribution of the intra-cluster medium is significantly flatter than the dark matter distribution, in particular for low-temperature clusters \citep[e.g.,][]{1999MNRAS.305..631A, 1999MNRAS.305..834E, 1999ApJ...517..627M}, due to non-gravitational heating.

\begin{figure}
\resizebox{\hsize}{!}{\includegraphics{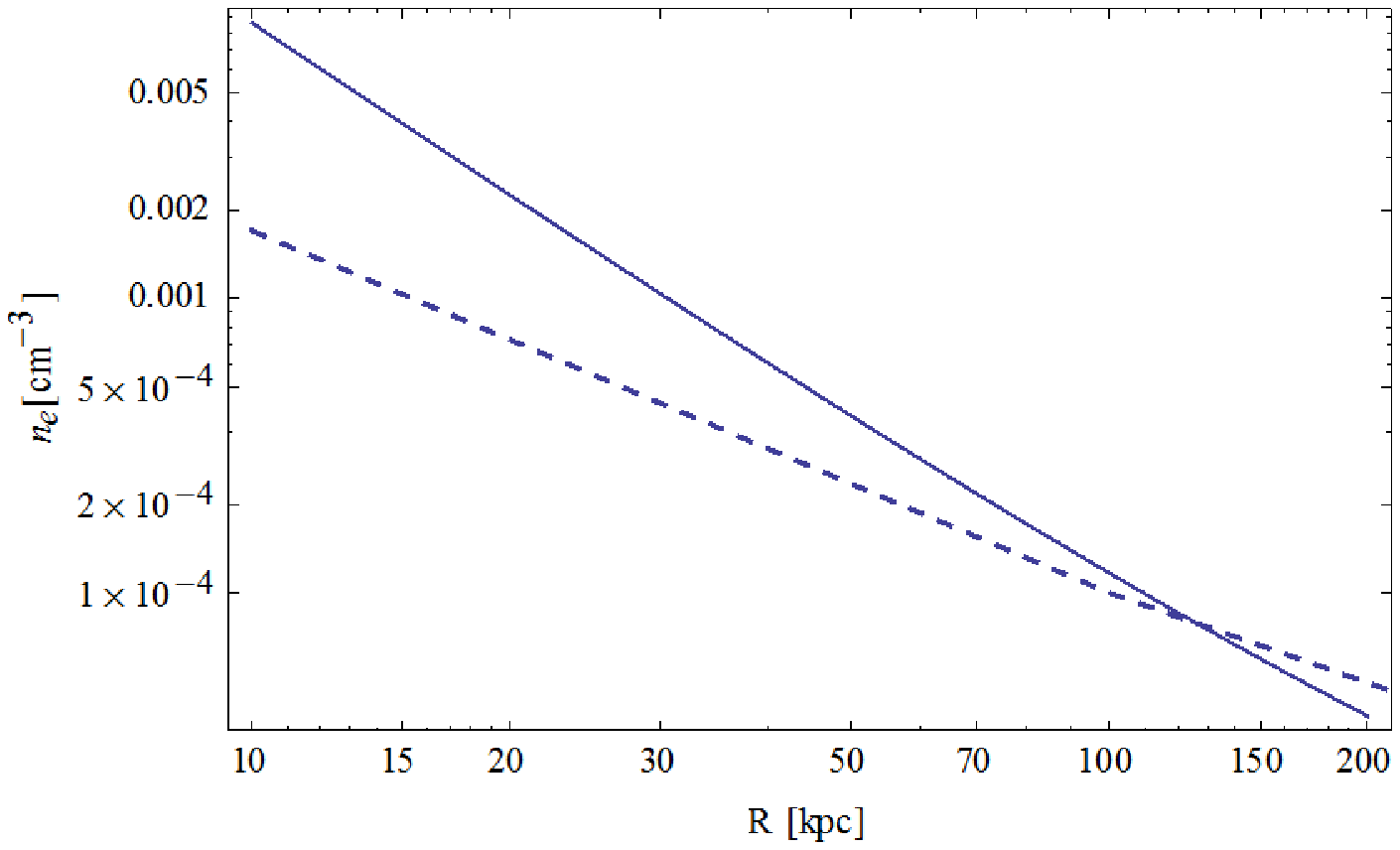}}
\caption{Electron number density for our coronal model. For comparison the dashed line  shows the electron  number density model taken from literature \citep[their 'model 2']{2006ApJ...639..590F}.
}
\label{Corona}
\end{figure}

In our modelling, we  assume that the density distribution of this gas settles down in hydrostatic equilibrium under the effect of the gravity of the dark matter component of the host galaxy (the MW in our case) with a temperature around $10^{6 \div 7} K$  (or $ \sim 0.086 \div 0.86$
 keV as common in plasma physics). The ion component provides the density and it is self-consistently added to the galactic potential described by \citet{2011A&A...525A..99P} to which  the reader should refer for a the detailed description. Following \citet{2006ApJ...639..590F} \citep[see also,]{2010ApJ...714..320A}, we adopt $Y = 0.25$ for the primordial abundance of helium, with a sound speed $v_s  = \sqrt {\gamma Zk_B T_e /m_i }  \cong 600\;{\rm{km}}\;{\rm{s}}^{ - 1} $ (with $T_e $ as the temperature in eV, $m_i $ as the ion mass, $Z$ as the charge state, $k_B $ as the
 Boltzmann constant and $\gamma $ as the adiabatic index \citep[see also][]{1994sse..book.....K}. The resulting model for the electron density  is  shown in Fig \ref{Corona}. Other approaches based on polytropic equations of state cannot add any significant insight to the problem due to the lack of observational constraints on the plasma temperature's spatial gradient or rotation in the range of distances we are interested to sample.

%%%%%%%%%%%%%%%%%%%%%%%%%Figure 1

\subsection{Tidal approximation for a pressure equation in non inertial reference systems.}\label{Tidal approximation for a pressure equation in non-inertial reference systems}

We limit ourselves to the simple case of a spherical dwarf galaxy whose size is given by a typical scale radius, $r_s $, e.g.,  the tidal radius or the effective radius, or any other radial scale length that might be obtained from observational data. If the dwarf galaxy is not a sphere, then the direction of motion must also be specified in its orbit around a massive companion in relation to the shape of the dwarf galaxy, and more sophisticated formulations are required.

We want to develop a formalism for the pressure acting on a dwarf galaxy while orbiting through the hot coronal gas of the MW (Section \ref{Galaxy model and Plasma distribution}). Quantifying the pressure is necessary in order to investigate the effects on the star formation rate as explained in Section \ref{Molecular clouds evolution and non gravitational heating}.

The pressure can be obtained from the Navier-Stokes equations in a potential-flow approximation once these equation are written in a non-inertial reference system, $S_1 $, comoving with the dwarf galaxy. This approach is based on the classical assumption that the interstellar medium (ISM) of a dwarf galaxy is much denser than the hot intergalactic medium (HIGM) of the MW. If we consider that the HIGM has a temperature greater than 10/100 times the dwarf galaxy ISM temperature (see Section \ref{Galaxy model and Plasma distribution}) then we can neglect the effect of the dwarf potential on the HIGM. Moreover the ISM will be considerably denser than the HIGM so that, to a first approximation, the flow of the hot gas is like that around a rigid body \citep[e.g.,][and references therein]{1982MNRAS.198.1007N}.
 In this case, we can proceed by adopting the standard formulation for the potential flow
\begin{equation}\label{potflow01}
	\varphi_{{\bm{v}}_0 }  =  - \left\langle {{\bm{v}},{\bm{\xi }}} \right\rangle \left( {1 + \frac{1}{2}\frac{{r_s^3 }}{{\left\| {\bm{\xi }} \right\|^3 }}} \right)
\end{equation}
\citep[e.g.,][]{ 1979mitf.book.....C, 2000ifd..book.....B}. Here ${\left\langle { \bm{a} , \bm{b} } \right\rangle }$ is the inner product between two arbitrary vectors $\bm{a}$ and $\bm{b}$, the subscript ${{\bm{v}}_0 }$ refers to the fluid velocity of the potential flow given by ${\bm{v}}_0  = \nabla \varphi_{{\bm{v}}_0 }$, ${\bm{v}}$ is the dwarf galaxy velocity in an inertial reference system $S_0$ centred on the MW, and ${\bm{\xi }}$ the vector radius in the non-inertial reference system $S_1$ whose origin is in the barycentre of the dwarf galaxy ${O'}$. Hereafter the prime-notation refers to the non-inertial reference system $S_1$.
The integration of the Euler equation is a standard procedure \citep[e.g.,][]{1959flme.book.....L, 2000ifd..book.....B} that leads to the classical Bernoulli equation that we would need to solve for the pressure $P$, but instead, we write it in a non-inertial reference system by adding the extra term ${\left\langle {{\bm{A}},{\bm{\xi }}} \right\rangle }$ that takes into account the non-inertial character of the reference system $S_1$ (where ${\bm{A}}$ is the acceleration of a fluid element due to the motion of the galaxy).
We get:
\begin{equation}
\frac{{\partial \varphi'_{{\bm{v'}}_0 } }}{{\partial t}} + \frac{P}{\rho } + \frac{{\left\| {{\bm{v}}'_0 } \right\|^2 }}{2} = f\left( t \right) - \Phi _{g}  - \left\langle {{\bm{A}},{\bm{\xi }}} \right\rangle.
\label{eq6}
\end{equation}
In this equation, $\rho $ is the coronal gas density distribution and $\Phi _{g}$ the gravitational potential acting on the fluid elements. We use a prime to indicate that the potential flow refers now to a non-inertial reference frame $S_1$ as well as its velocity ${{\bm{v}}_0 }$. This equation closely resembles the classical Bernoulli equation but for the presence of the non-inertial term \citep[see Section 3.1 of][for a more extended discussion]{2009A&A...499..385P}.
Finally, we fix the arbitrary $f\left( t \right)$ (constant over all the streamlines for inviscid irrotational fluids) by imposing the boundary condition that the hot gas corona is in hydrostatic equilibrium. We easily obtain
\begin{equation}
f\left( t \right) = \frac{{\left\| {\bm{v}} \right\|^2 }}{2}.
\label{eq7}
\end{equation}

The last approximation we want to introduce before solving Eqn. \eqref{eq6} for the pressure $P$ is the tidal approximation. Because we are interested in the pressure acting in the regions close to the dwarf galaxy, we perform a Taylor expansion of the total gravitational potential $\Phi _g \left( {\bm{x}} \right)$ acting on the generic element of the fluid located at ${\bm{x}}$ close to the barycentre (located at ${\bm{R}_0}$) of the dwarf galaxy:
\begin{equation}\label{Taylor}
\Phi _g \left( {\bm{x}} \right) \simeq \Phi _g \left( {{\bm{R}}_0 } \right) + \nabla \Phi _g \left( {{\bm{R}}_0 } \right)\left( {{\bm{x}} - {\bm{R}}_0 } \right) + ...
\end{equation}
so that ${{\bm{x}} - {\bm{R}}_0 }={\bm{O}\bm{\xi }}$ with ${\bm{O}} \in SO\left( 3 \right)$ is the generic rotation matrix between the orthonormal basis of the inertial reference system $S_0$ and $S_1$ with transpose $\bm{O^\text{T}}$, and the approximation holds for ${{\bm{\xi }} \ll {\bm{R}}_0 }$  and ${{\bm{\xi }} \ll {\bm{x}} }$. Considering Eqn. \eqref{Taylor} and that ${{\bm{A}} =  - \nabla \Phi _g }$, we easily obtain from Eqn. \eqref{eq6} the required equation for the pressure in the tidal approximation
\begin{equation}
P = \rho \left[ {\frac{{v^2 }}{8}\left( {4 - 9\sin ^2 \vartheta } \right) - \Phi_g \left( {\bm{R}_0} \right) + r_s^2 \left\langle {{\bm{O^\text{T} T O \hat n}},{\bm{ \hat n}}} \right\rangle } \right],
\label{eq9}
\end{equation}
where ${v^2 = \left\| {\bm{v}} \right\|^2 }$ is the square of the norm of the velocity vector of the dwarf galaxy, $\vartheta $ is the angle between the direction of the velocity vector and the normal ${{\bm{\hat n}}}$  \citep[e.g.,][]{1959flme.book.....L}, and the equation is evaluated at the scale radius: ${{\bm{\xi }} = r_s {\bm{\hat n}}}$ with  ${\bm{\hat n}} = \frac{{\bm{\xi }}}{{\left\| {\bm{\xi }} \right\|}}$. Finally, we have introduced the tidal tensor ${{\bm{T}} =  - {\bm{H}}\left( {\Phi _g \left( {{\bm{R}}_0 } \right)} \right)}$ defined as the negative to the Hessian matrix ${\bm{H}}$ in the Taylor expansion of Eqn. \eqref{Taylor} stopped at the first order \textit{in the potential }and evaluated at the dwarf galaxy barycentre position.

Equation \eqref{eq6} is completely determined once the orientation of the reference system $S_1$ with respect to $S_0$ is defined, i.e. when we define the rotation matrix $\bm{O}$. We use the Cayley-Klein parameters \citep[e.g.,][]{2002clme.book.....G} to orient $S_1 $ with the Frenet-Serret reference frame $\left\{ {{\bm{\hat e}}_T ,{\bm{\hat e}}_N ,{\bm{\hat e}}_B } \right\}$ once the orbit is numerically integrated where the tangent vector ${\bm{\hat e}}_T $, the normal ${\bm{\hat e}}_N $ and the binormal ${\bm{\hat e}}_B $ are its instantaneous associated unitary vectors. From the orbit we can infer the kinetic pressure $P = P\left( {{\bm{x}},t} \right)$, where ${\bm{x}} = {\bm{x}}\left( t \right)$ is the solution of the EoM in the inertial reference system $S_0 $ centred on the host galaxy defined by the MW gravitational potential $\Phi _{g} $ as in, e.g., \citet{2011A&A...525A..99P} at $t_{\text{lb}}  = 0$, i.e. the present-day potential (with $t_{\text{lb}}$ look-back time). For simplicity in the following we are neglecting the  time dependence of the gravitational potential.

We highlight a few characteristics of Eqn. \eqref{eq9}:
\begin{itemize}
	\item It contains an explicit dependence on the size of the system $r_s $. This parameter can be easily changed from the tidal radius to the half-mass radius, or  can be related to  the light profile depending on the specific applications.
 	\item It shows a dependence on the direction  ${\bm{\hat n}}$ along which we want to estimate the pressure.
\item It reduces to the standard results in the special case of the absence of a gravitational field, in the case of an inertial system of reference and for circular orbits.
\end{itemize}

\begin{figure}
\resizebox{\hsize}{!}{\includegraphics{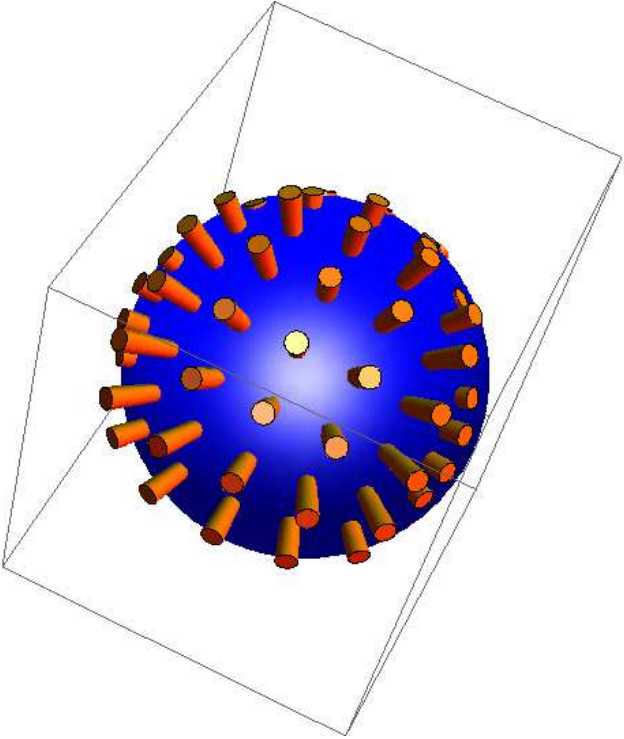}}
\caption{Plot of the pressure acting on the ''surface'' of a dwarf galaxy (blue sphere) with orbit E050 at the look-back time $t_{lb}=9$ Gyr (see the text for further details and Fig. \ref{Panel2}). Orange pistons represent the pressure intensity on the dwarf, the higher the piston the higher is the pressure. %A discussion on the angular limit of validity is seen in Fig. \ref{PasettEqPress}.
}
\label{Experimental}
\end{figure}

In Figure \ref{Experimental} we show the pressure in the reference system $S_1$ for a dwarf galaxy orbiting with an eccentricity $e=0.5$ and pericenter $r_p=200$ kpc in the MW external potential. This model of the galaxy is later referred to as model E050 (see further details in Fig. \ref{Panel2}). Along  each direction perpendicular to an ideal  surface enclosing the ISM of  the dwarf galaxy (indicated by the blue color) a cylinder or piston is plotted. Its height is proportional to the intensity of the pressure  normalized to the value at the stagnation point (the ideal point where the flow velocity vanishes). Due to the complexity of the terms in equation \eqref{eq9} the behaviour of the cylinder is clearly not linear.

%%%%%%%%%%%%%%%%%%%%%%Figure 2

Finally, we can further refine our model in two way:
\begin{itemize}
	\item By considering the variability of the scale radius that can occur during the evolution of the dwarf galaxy in its tidal interacting with  the host system. In this case, due to the linearity of the Laplace equation for the potential flow, we can simply add the potential flow of a reducing size dwarf
\begin{equation}\label{potflow02}
	\varphi _{{\bm{v}}_0 }  =  - \frac{{\dot r_s r_s^2 }}{{\left\| {\bm{\xi }} \right\|}}	
\end{equation}
 to the previous potential flow of Eqn. \eqref{potflow01}. In this case to deduce the pressure term is more tedious but straightforward and we omit it. 
 \item By considering that the orbits of dwarf galaxies are different in nature according to whether  their velocity is subsonic or transonic/supersonic. When the galaxy velocity  becomes comparable with or exceeds that of the sound in the inter galactic medium, effects due to the compressibility of the fluid become  important and shock waves can occur within the galaxy or backward throughout the ISM \citep[e.g.,][]{1959flme.book.....L}. In the literature these concepts find  application in problems related to the free fall of galaxies in rich clusters of galaxies \citep[e.g.,][their Eqn. 6]{1996ApJ...459...82H}. Nevertheless we will not encounter this situation in the dwarf galaxies orbiting through the hot MW coronal gas that we are going to analyse since the velocity and density involved are far from the supersonic regime. We refer the interested reader to Appendix \ref{pressureprofiles} for the test of the pressure equation and its relation with the standard supersonic formalism.
\end{itemize}

%%%%%%%%%%%%%%%%%%%%%%%%%%Figure 3
\begin{figure*}
\centering
\resizebox{\hsize}{!}{\includegraphics{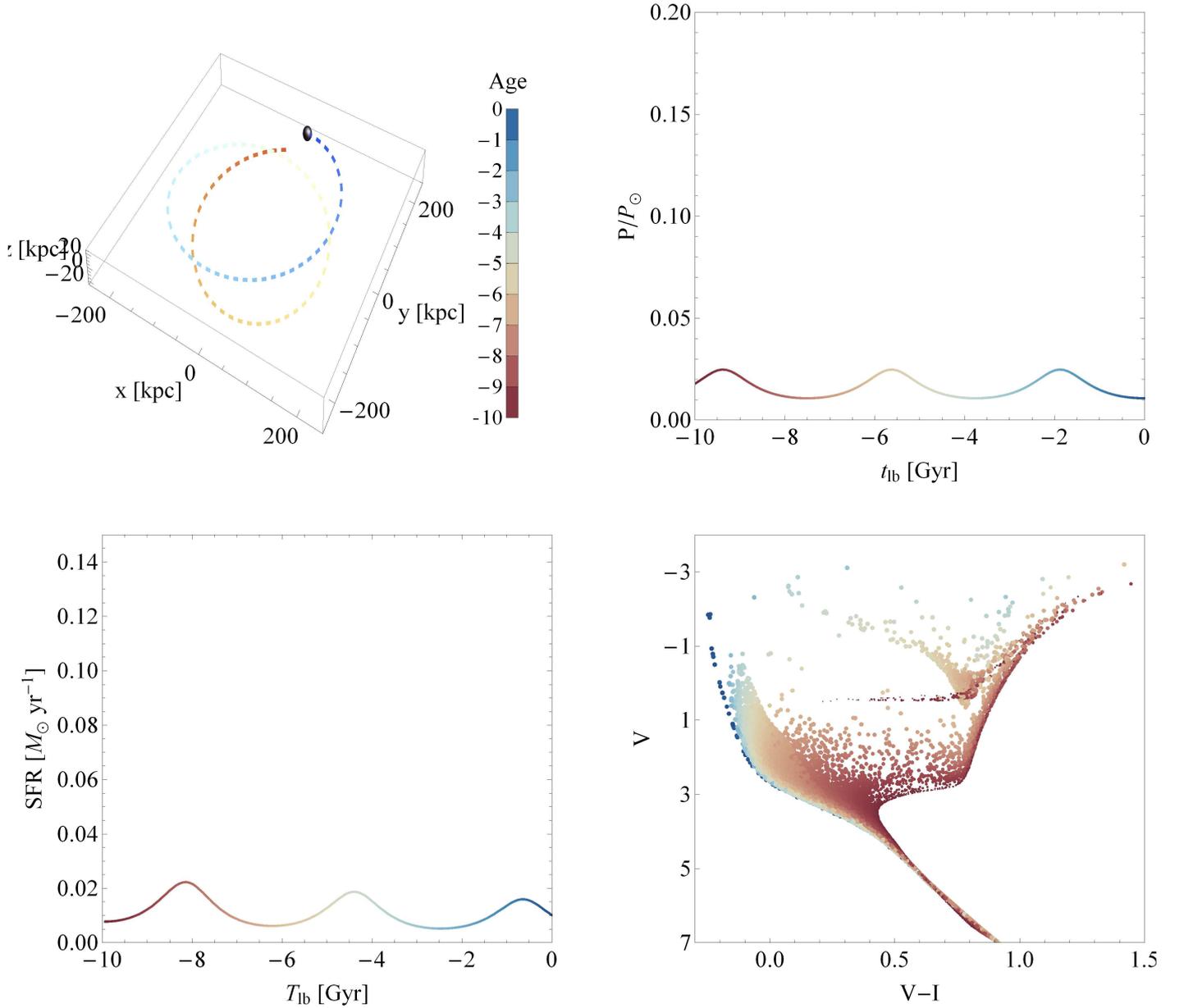}}
 \caption{For a test satellite orbit with eccentricity e=25 and baryonic mass $M_b  \cong 10^7 M_ \odot$ we show  the path (upper left panel), the resulting pressure profile (upper right panel) normalized to $P_ \odot  $ defined in the text, the star formation rate (lower left panel) and the resulting CMD (lower right panel). The  colour bar is plotted only in the upper left panel, red for the  oldest stars at the beginning of the orbital path $t = 0$, blue for the youngest stars born after 10 Gyr of orbital evolution. In the pressure panel, $t_{lb}$ is the look-back time in Gyr. In the CMD the size of the diamonds is proportional to the metallicity Z of the stars as
 obtained the Padova CMD simulator YZVAR fed with  the stellar  models of \citet{2009A&A...508..355B} (metal rich stars are indicated by large diamonds).
}
 \label{Panel1}
\end{figure*}

%%%%%%%%%%%%%%%%%%%%%%%%%%%%Figure 4
\begin{figure*}
\centering
  \resizebox{\hsize}{!}{\includegraphics{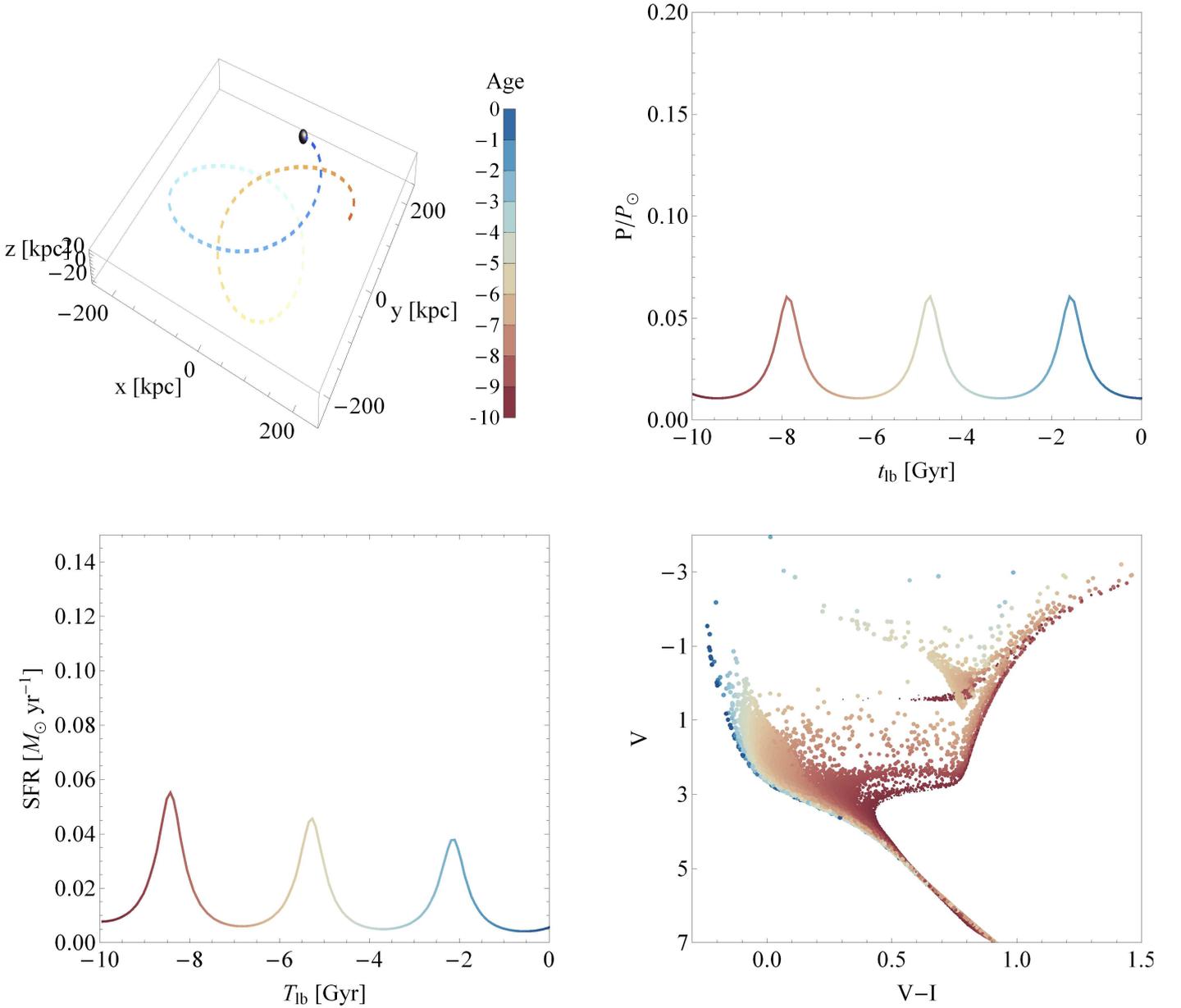}}
 \caption{As in Fig. \ref{Panel1}  but for an orbital eccentricity e=0.5 of the satellites. The pressure panel refers to the stagnation point. $t_{lb}$ is the look-back time in Gyr. For $t_{lb}=9$ Gyr the surface pressure is represented in Fig. \ref{Experimental}.
}
 \label{Panel2}
\end{figure*}

%%%%%%%%%%%%%%%%%%%%%%%%%%%%%%Fig 5
\begin{figure*}
\centering
  \resizebox{\hsize}{!}{\includegraphics{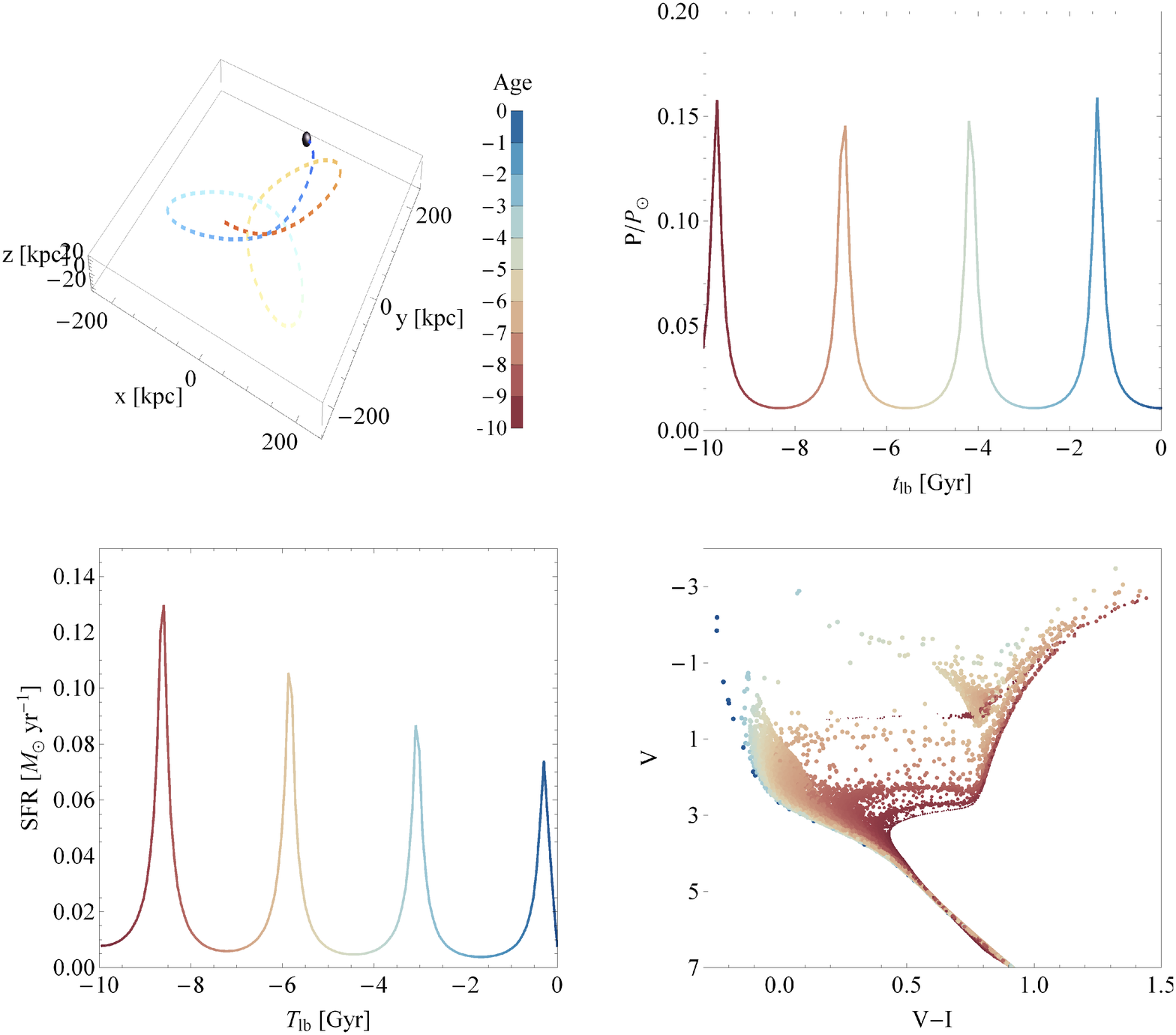}}
 \caption{The same as in Fig. \ref{Panel2} but  for the eccentricity e=0.75.
}
 \label{Panel3}
\end{figure*}

In the panels of Fig. \ref{Panel1}, \ref{Panel2} and \ref{Panel3} we plot three representative orbits and the relative pressure computed with Eqn. %(1.8)
\eqref{eq9} along the direction of motion ${\bm{\hat e}}_T $. The pressure is plotted for the unitary scale parameter $r_s  = 1$, normalized pressure and time in Gyr. As arbitrary normalization value for the pressure we have chosen the standard literature value for pressure from the random motion of molecular clouds in the solar neighbourhood, $P_{\odot} = 3 \times 10^4 k_B \;\text{cm}^{ - 3} \text{K}$ with  $k_B$ the Boltzmann constant  \citep[e.g.,][and Appendix B]{1997ApJ...480..235E,1998ApJ...509..587F}.
As expected the maximum pressure experienced by the dwarf is at the point of closest passage in the orbit mostly as a result of the higher velocity of the dwarf.
The orbit selection is based on the eccentricity. We assume a starting distance of 200 kpc in the MW potential described by \citet{2011A&A...525A..99P},  then we search for initial conditions of the  orbit in the MW plane supposed to be at rest in $S_0 $
 starting from ${\bm{w}}_0  = \left\{ {{\bm{x}}_0 ,{\bm{v}}_0 } \right\}|e = \left\{ {0.25,0.50,0.75} \right\}$
 with initial conditions  ${\bm{x}}_0  = \left\{ {0,200,0} \right\}$
 kpc, $e \equiv  {{r_a  - r_p } \mathord{\left/
 {\vphantom {{r_a  - r_p } {r_a  + r_p }}} \right.
 \kern-\nulldelimiterspace} {r_a  + r_p }}$
  where  $r_a ,r_p $ are apo-centre and peri-centre  of the orbits respectively. We will refer to the 3 orbits evolved in the MW model introduced in the previous Section by their  eccentricity E025, E050, E075.

\subsection{External gas consumption effect: ram pressure stripping, Kelvin-Helmholtz and Rayleigh-Taylor instability }
\label{External gas consumption effect}

Given the prescription for the pressure along an assigned orbit, we can reconstruct the SFH of the galaxy by resolving the system of equations \eqref{eq1}. However, we need also to consider  the effects of external instabilities  that can affect  the gas consumption processes and, as a consequence,  the SFH of a dwarf galaxy as well.
We consider two distinct types of stripping processes: (i) instantaneous stripping due to ram pressure and Rayleigh-Taylor instabilities and (ii)  continuous stripping  due to the Kelvin-Helmholtz instability.

The approach adopted  to deal with the kinds of stripping is  different. For the ram pressure and Rayleigh-Taylor instabilities we extend the classical recipe by making use of the formalism developed in the previous section. It has the advantage of taking  the galaxy size and the direction relative to the motion into account analytically. In this way, we generalize the literature results, see e.g., the classical study by \citet{1972ApJ...176....1G} and the recent works by \citet{2005A&A...433..875R}  for disk galaxies or those by \citet{2000ApJ...538..559M} on the thermal pressure in the galactic centre.

Stripping of gas by ram pressure occurs  when the pressure times the unit area, i.e. $ - P_{{\rm{gas}}} {\bm{\hat n}}dA$, is larger than the restoring  gravitational force for a unit mass of gas (HI or molecular, see e.g., \citet{2009ApJ...693..216K}) of the dwarf galaxy clouds (with $\Phi _{\text{dg}}$, $M _{\text{dg}}$ and $\rho_{\text{dc}}$ the potential, mass and density respectively in $S_1$ of the dwarf galaxy):
\begin{equation}\label{GravF}
\begin{gathered}
  {\mathbf{g}}\left( r \right) =- \frac{{d\Phi _{{\text{dg}}} }}
{{d\xi }}{{\bm{\hat n}}} =  - \frac{{GM_{{\text{dg}}} \left( {\xi ,t} \right)}}
{{\xi ^2 }}{{\bm{\hat n}}} \hfill \\
  M_{{\text{dg}}} \left( {\xi ,t} \right) = 4\pi \int_0^{r_s } {\rho _{{\text{dg}}} \left( {\xi ,t} \right)\xi ^2 d\xi }.  \hfill \\
\end{gathered}
\end{equation}
For simplicity, the time dependence is only in the total mass derived from  Eqns. \eqref{eq1} whereas   the scale of the structural parameters remains unchanged.

This formalism can be easily applied to study the Rayleigh-Taylor instability that in order to occur  requires
\begin{equation}\label{RTcond}
\frac{{GM_{{\text{dc}}} \left( {\xi ,t} \right)}}
{{\xi ^2 }}M_{{\text{gas}}} \left( {\xi ,t} \right) < \frac{1}
{2}c_d P,
\end{equation}
where the drag coefficient, $c_d $, is taken to be equal to one for simplicity \citep[e.g.,][]{2008A&A...483..121R}. It is also interesting to note that the formulation adopted in Eqn. \eqref{eq9} is ``local'', in the sense that it contains information regarding the direction $\bm{\hat n}$ of the point where the condition Eqn. \eqref{RTcond} holds. In this way we can  take into account of the direction of observation ${{\mathbf{\hat n}}}$.
We see here in analytical way, which dwarf galaxy parameters are relevant for the Rayleigh-Taylor instability. This effect has a dependence on the velocity of the dwarf $v$, the scale length $r_s$ and the direction of the motion trough the function $\sin \vartheta $.
As soon as we look away from the direction of the velocity vector (i.e. $\vartheta$ increases) there are zones of the dwarf galaxy that are more stable against this type of instability i.e., the pressure in Eqn. \eqref{RTcond} decreases. At the same way we see that the more massive the galaxy is, the less efficient is Rayleigh-Taylor stripping.

Finally, the gas surviving the instantaneous ram pressure stripping can be later removed by the Kelvin-Helmholtz instabilities occurring at the interfaces between the ISM in the dwarf galaxy and the HIGM, even if their effects are partially suppressed by  gravity. Since we are not interested in wakes we can safely neglect the ICM viscosity  \citep{2008MNRAS.388L..89R} and consider this stripping process as continuous and  mostly affecting the gas in the flow past the dwarf galaxy \citep{1982MNRAS.198.1007N}. This effect is expected to decrease the total mass of the dwarf galaxy in the  Eqns. \eqref{eq1} according to
\begin{equation}
\left. {\frac{{dM}}
{{dt}}} \right|_{{\text{KH}}}  = \pi r_s^2 \rho v,
\end{equation}
with a natural time-scale $\tau _{{\text{KH}}}  = {\raise0.5ex\hbox{$\scriptstyle {M_{gas} }$}
\kern-0.1em/\kern-0.15em
\lower0.25ex\hbox{$\scriptstyle {\left. {\frac{{dM}}
{{dt}}} \right|_{{\text{KH}}} }$}}$ and an instability condition that comes from the classical perturbative approach \citep[e.g.,][]{1961hhs..book.....C}
\begin{equation}
k > \frac{g}
{{v^2 }}\left( {\frac{1}
{{{\raise0.7ex\hbox{${\rho _{{\text{HICM}}} }$} \!\mathord{\left/
 {\vphantom {{\rho _{{\text{HICM}}} } {\rho _{{\text{ICM}}} }}}\right.\kern-\nulldelimiterspace}
\!\lower0.7ex\hbox{${\rho _{{\text{ISM}}} }$}}}} - {\raise0.7ex\hbox{${\rho _{{\text{HICM}}} }$} \!\mathord{\left/
 {\vphantom {{\rho _{{\text{HICM}}} } {\rho _{{\text{ICM}}} }}}\right.\kern-\nulldelimiterspace}
\!\lower0.7ex\hbox{${\rho _{{\text{ISM}}} }$}}} \right)
\end{equation}
with $k$ wavenumber of the dominant wavelength gas ablation for Kelvin-Helmholtz instability assumed to be of the order of the scale parameter $r_s$ \citep[e.g.,][]{1993ApJ...407..588M} and $g = \left\| {\mathbf{g}} \right\|$ from Eqn. \ref{GravF}.

\section{Synthetic colour magnitude diagrams}
\label{Synthetic colour magnitude diagrams}

In order to explore the connection between the SFH of dwarf galaxies and their physical causes which may be either internal (i.e. star formation itself  and feedback) or external (environment processes: i.e. orbits, tidal field and gas dissipative phenomena), we have to deal with the constraints imposed by the observations. The most direct detection of the SFH is  based on the study of CMDs with the aid of the synthetic CMDs technique. In brief, given a  star formation rate (SFR),  an initial mass function (IMF), a binary fraction and a chemical enrichment law (expressed by the helium to heavy element ratio $\Delta Y /\Delta Z$), one can generate  synthetic  CMDs in any photometric system  to be compared with their observational counterparts. To this aim, we make use of the Padova synthetic CMD generator \citep[e.g.,][]{2002A&A...392.1129N}. Originally this method was developed for testing stellar evolution models of stars of different mass \citep[e.g.,][]{1985A&A...150...33B} or stellar clusters  \citep{1989A&A...219..167C,1995A&A...301..381B}. We use here the most recent version of this tool that utilizes of the library of stellar models calculated by   \citet{2009A&A...508..355B}. We refer to the rich literature developed by the Padova's astrophysics group for further details on YZVAR, the implementation techniques and the stellar models in use.

In the lower right panels of  Figs. \ref{Panel1}, \ref{Panel2}, \ref{Panel3} we show synthetic CMDs in the V-I pass-bands  without correction for the distance modulus. To study the effects of different orbits on the SFH and CMD in turn, we adopt always the same model for the dwarf galaxy, evolved along orbits with different eccentricities and leave all other parameters unchanged.

By comparing the CMD for the models (E025, E050, and E075) it is evident that SFH and associated CMDs depend on the orbit eccentricity.
The CMD of the E025 model shows continuity between the episodes of star formation, which correspond to continuous gas consumption. This correlates with the orbital energy of the dwarf galaxy in two different ways. On  one hand, we have a dependence on the location of the dwarf, i.e. on the configuration space. The pericentre of the dwarf galaxy is less close to the host galaxy than for the other simulations with higher eccentricity (cf. Fig. \ref{Panel2} and \ref{Panel3}). According to  Eqn. \eqref{eq9} this reduces the influence of the external tidal field and coronal gas via  the potential $\Phi $ and density $\rho $. On the other hand, there is a  quadratic dependence on the velocity $ \propto v^2 $, with a maximum at the pericentre passage that enhances the  ram pressure stripping. Therefore, the MW dwarf galaxies with  smaller eccentricity (say up to $e = 0.25$) are expected not to show clear signatures in the gas consumption  of effects of dynamical nature in their CMDs.

Different results are  obtained for higher eccentricities, cf. the CMD of  Fig \ref{Panel3}. In this case, we see different subgiant branches as signatures  of an irregular consumption of the gas. Up to three to four episodes of SF are evident. The  old metal-poor stars (cyan colour code and small size symbols)  are clearly separated from  from the young  and metal rich ones (red colour code  and larger symbols). Despite the overlap of the main sequences (MSs) that should be analysed with finer resolution to resolve the sub-components,  the three to four sub-giant branches (SGBs) are clearly indicating three to four episodes.  The lowest and faintest SGB corresponds to stars of about $0.8 M_{\odot}$ and 9 Gyr among which older and lighter stars are also present. This is followed by several episodes, up to ages of about  $ \sim 3.5/4.0$ Gyr and masses of about $m \simeq 1.0M_ \odot  $. Episodes younger than this are difficult to disentangle. It is beyond the aims of this study to analyse the CMD in a very detailed fashion. Our main purpose here is to show that dynamical effects can in principle leave traces of their occurrence in the CMDs of the stellar content of a dwarf galaxy,  see \citet[]{2011A&A...530A..59C} and Section \ref{Conclusions} below).

\section{Macrophysics vs. microphysics. An application to the Local Group dwarf galaxy Carina} \label{An application to the local group}
In the literature there are different approaches to the study of the dissipative phenomena in the astrophysics context. The motivation of these studies is the attempt to explain observational evidence with perturbative approaches to the Navier-Stokes equations \citep[for a review see, e.g.,][]{1961hhs..book.....C}.
In relation to the astrophysics of galaxies, these works have been developed in connection with integrated properties (e.g., total mass, central velocity dispersion etc.) and star formation processes \citep[e.g.,][]{1980ApJ...240L..83L, 1982MNRAS.198.1007N}. Recently, thanks to the continuous development of the numerical techniques, we see great efforts in the study of these astrophysical processes from the microscopic point of view \citep[e.g.,][]{2000ApJ...538..559M,2004MNRAS.352.1426Q}.

We present in Appendix A a few tests on the compatibility of the results obtained with our macrospcopic approach based on integrated equations and the detailed description we can obtain from codes available in the literature that solve the microphysics of the instability processes.
Hereafter, we show the utility of our technique in a real astrophysical context, because the ultimate achievement of this technique is indeed to understand the relation between the dissipative phenomena and their connection with observable quantities, the orbits of dwarf galaxies and their CMD.

The most natural laboratory where to search for real observational data is surely the LG of galaxies. The LG of galaxies are of paramount importance because of their proximity and the wealth of data available so that studies of the past SFH are possible  \citep[e.g.,][]{1997RvMA...10...29G,2008ApJ...686.1030O}. In the recent past, the acquisition of large samples  of spectroscopic data made it possible  to measure the internal kinematics of some of these dwarf galaxies as well as, in a number of cases, their stellar metallicities and abundances \citep[e.g.,][]{2008AJ....135.1580K,2006AJ....131..895K,2007AJ....133..270K} thus initiating a new generation of full chemo-dynamical models able to take both orbital parameters deduced from proper motions and internal kinematics into account. E.g., concerning the Magellanic Clouds there is a rich literature on the determination of their proper motions \citet[e.g.][]{2006ApJ...652.1213K}. In a recent paper, \citet{2011A&A...525A..99P} presented a study of the Carina dwarf spheroidal galaxy in which, taking into account most of the available observational constraints, we recovered in a self-consistent fashion both the past SFH and the internal kinematics of this dwarf galaxy.

\begin{figure}
\resizebox{\hsize}{!}{\includegraphics{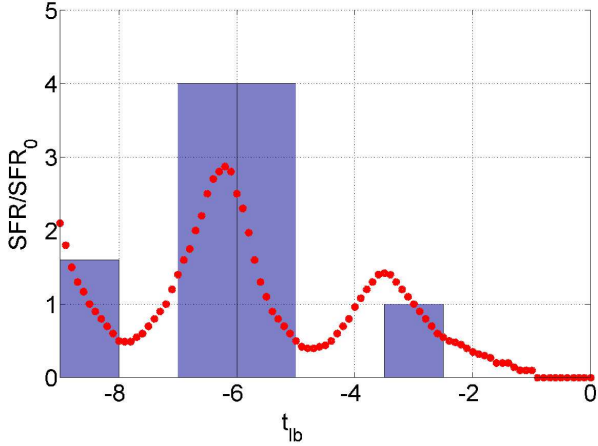}}
\caption{Recovered episodic star formation history of the Carina dwarf galaxy superposed on  the normalized histogram bars from the data of \citet{2003ApJ...589L..85R} (see also \citet{2011A&A...525A..99P}).}
\label{SFRc}
\end{figure}

We use the results of \citet{2011A&A...525A..99P} in the light of the new technique we have here developed. In this paper we recover for the first time self consistently the star formation history of Carina dwarf galaxy as consequence of the pericenter passages which trigger SF. The technique that we adopted is based on a chemo-dynamical code \citep[e.g.,][]{2003Ap&SS.284..865B} with new star formation recipes described in detail in \citet{2010A&A...514A..47P}. Our simulations reach a mass resolution of $118M_ \odot  $ resolving the core of the dwarf galaxy with about 127000 particles (stars) over about 700000 total particles.
In this approach the bursts of star formation in the evolution of Carina are the result of microscopic physical effects: the orbits of particles are integrated and some recipes rule the star formation and feedback that each particle experiences or exerts on the others. We show here that our earlier solution for the star formation history of Carina is \textit{not} unique and it can be easily achieved by our new methodology.
We proceed with a few simpler assumptions than in the work of  \citet{2011A&A...525A..99P}, because this exercise is aimed to be a qualitative example and not a quantitative study of the importance of the relative dissipative effects (that is left for a forthcoming work, Pasetto et al 2012 in preparation). We do not span the entire range of orbital parameters as in \citet{2011A&A...525A..99P}, moreover the starting model adopted for the primordial dwarf galaxy is inspired by the results \citet{2011A&A...525A..99P}, except that for the sake of simplicity we neglect here the temporal evolution of the model for the MW. The MW potential is described by the following gravitational potential
\[
\begin{gathered}
  \Phi _{{\text{halo}}}^{{\text{MW}}} \left( {R,z} \right) = k_1^2 \log \left( {R^2  + k_2^2  + \frac{{z^2 }}
{{k_3^2 }}} \right) \hfill \\
  \Phi _{{\text{disk}}}^{{\text{MW}}} \left( {R,z} \right) =  - \frac{{GM_{{\text{disk}}} }}
{{\sqrt {R^2  + \left( {k_4  + \sqrt {z^2  + k_5^2 } } \right)} }} \hfill \\
  \Phi _{{\text{bulge}}}^{{\text{MW}}} \left( {R,z} \right) =  - \frac{{GM_\text{bulge} }}
{{\sqrt {R^2  + z^2 }  + k_6 }} \hfill \\
\end{gathered}
\]
with $G$ being the gravitational constant and the family of parameters $\left\{ {k_i } \right\}, i = 2,..,6$, given by
$\left\{ 12,0.8,6.5,0.26,0.7 \right\}$ in kpc, and $k_1=130.8$ in km s$^{-1}$, $M_\text{bulge}=3.4 \times 10^{10} M_ \odot  $ and $M_{\text{disk}}=1.0 \times 10^{10} M_ \odot$. Over this external field only the family of orbits in the neighbourhood of the phase-space point ${\mathbf{x}} = \left\{ {22,89, - 34} \right\}$ kpc, ${\mathbf{v}} = \left\{ { - 136,13, - 45} \right\}$ in a reference frame $S_0$ collinear with the MW potential are investigate and extracted from the family of proper motions in Table 1 of \citet{2008ApJ...680..287M} with the minimization action technique developed in \citet{2011A&A...525A..99P}.  We adopt an initial potential-density pair for Carina from the subfamily-$\left( {1,4,\gamma } \right)$ of the Zhao-models \citep{1996MNRAS.278..488Z}, widely known also as the $\gamma$-models. We assume $M\left( r \right) = M_{{\rm{bar}}} \left( {\frac{r}{{r + r_s }}} \right)^{3 - \gamma } $ with $M_{{\rm{bar}}}  = 4.8 \times 10^7 M_ \odot  $ being the baryonic mass of the dwarf galaxy, $r_s  = 0.7$ kpc and $\gamma  = 3/2$.  Finally, we adopt an initial SFR $\psi_0  = 0.029M_ \odot  yr^{ - 1} $ at $t_{\text{lb}}  = 10 $Gyr ($t_{\text{lb}}$ stands for look-back time). The corresponding initial value of the pressure required to balance the cloud destruction rate and the star formation rate is  $P = 0.01P_ \odot  $. The results for the recovered SFH  are plotted in Fig. \ref{SFRc}. Finally, stellar models in system Eqn. \eqref{eq1} and for the H-R diagrams in the low mass regime (from 0.15 to 2.5 $M_ \odot$) come from the work of \citet{2008A&A...484..815B}, and for higher mass from \cite{2009A&A...508..355B}. The technique is based on the the earlier works of \citet{1993A&AS..100..647B}, \citet{1994A&AS..105...29F,1994A&AS..105...39F}, \citet{1996A&AS..117..113G} but is updated in the opacities, rates of energy loss by plasma neutrinos and nuclear network integration \citep[see][and references therein]{2009A&A...508..355B}.

%%%%%%%%%%%%%%%%%%%%%%%%%Figure 6

By straight comparison of our plot in Fig. \ref{SFRc} with Fig. 11 in \citet{2011A&A...525A..99P}, the similarity is remarkable, thus confirming the original idea contained in that paper that the SFH of the Carina dwarf galaxy can be explained as results of pericentre passages. But the techniques are absolutely different: In our approach we do not integrate the orbits of the single particles. There are not microscopic particles but macroscopic analytical properties (total mass, density profiles, etc.). The star formation is the result of the pressure on the molecular cloud model, and the pressure depends on the orbit thought the MW potential, the tidal tensor as well as ram pressure, Kelvin-Helmholtz and Rayleigh-Taylor dissipative phenomena.
The major difference is that the effect of the tidal field of the MW is fully accounted for by solving the EoM  for the single dwarf galaxy particles in a N-body simulation, while here it is considered only by means of a first order perturbation, ultimately by the tidal tensor ${\mathbf{T}}$. More importance is given here to external instability processes that were not taken into account in \citet{2011A&A...525A..99P}. We have two different techniques that match the same observational constraint, i.e. the star formation history.
We point out another important difference with respect to the results obtained by  \citet{2011A&A...525A..99P}: while a high resolution chemo-dynamical simulation requires about 4 months on a special hardware machine, the result presented here needed only about 8 hours on a parallel cluster of processors. Moreover, the present technique can impose more straightforward constraints on the gas fraction adopted for modelling Carina that are impossible to reach because of particle resolution problems in a fully hydrodynamical NB-TSPH  simulation. In this respect the new  technique is superior to that followed by  \citet{2011A&A...525A..99P}.
The high speed  of the present algorithm  is partially due to the fully analytical formulation and partially to the high scalability of the code. We refer the interested reader to Appendix A for details on the parallelization performance of the code. Alternatively, the fully hydrodynamic code is best suited  to studies of the internal kinematics  that we have not taken into consideration here.

\section{Discussion, open questions, future developments and conclusion}
\label{Conclusions}

We have developed a method and companion code able to handle in a fast way the major phenomena affecting the SFH of extended dwarf galaxies moving through an external environment and compared  the results with those obtained from  a fully hydrodynamical code \citep{2011A&A...525A..99P}. This formalism stands on a novel interpretation of the classical results for a flow past an extended spherical object \citep[e.g.,][]{2000ifd..book.....B, 1959flme.book.....L} where we considered the problem in a non-inertial reference frame immersed in an external field and we exploited the tidal approximation for the external gravitational interaction \citep[e.g.,][]{2009A&A...499..385P}.

The advantage of this method, if compared to fully hydrodynamic codes, is its high speed (without imposing the restriction of a point-mass approximation) which is possible thanks to a new  direction-dependent formulation for the pressure effects described in a non inertial reference system and that is a valuable alternative to the classical works of \cite{1972ApJ...176....1G} or \citet{1982MNRAS.198.1007N}.

The technique is  very suitable to impose constraints on the orbits based on the SFH of  dwarf galaxies. We have shown a possible application to the case of  Carina to confirm the validity of the approach adopted. Another natural applications is the study of the triplet Large Magellanic Cloud, Small Magellanic Cloud and MW by taking into account their SFH. A further possible application is to constrain the phase-space parameters of the  external group of dwarf galaxies. The apparent lack of intermediate age populations in some early-dwarfs around Cen A \citep[see][]{2011A&A...530A..59C} or in other objects for which, due to their distance, only the most luminous part of the CMD can be resolved into stars could be  investigated using the  structural and dynamical constraints we have proposed here. In  studies of such more distant dwarfs, spectroscopic measurements of internal kinematics are still difficult to obtain and often only  photometric data are available. As a consequence, the investigation of these important issues with complete chemo-dynamical codes is time-consuming and  time consuming, whereas a fast and flexible technique and companion codes have better chances.

Finally, while the competition of the different dissipative phenomena to produce an observable quantity is surely a target of this project, it is not the primary goal. A dwarf galaxy is affected in its orbital evolution by several phenomena that all act jointly to produce an observational result, making it difficult to constrain the contribution of one individual mechanism. Our technique is, by itself, not able to disentangle the relevance of all those different phenomena because, as just seen in the exercise for Carina dwarf galaxy, the same result can be achieved with different methods. Nevertheless, this approach is a step forward in the inclusion of the stellar population algorithms to add a further observational constraint to the orbit investigation.

Of course, the simplicity of the adopted approach comes also  with a few technical limitations and drawbacks that should be amended.
For instance, in the Rayleigh-Taylor instability the magnetic field can play a role \citep{1961hhs..book.....C}. Moreover the magnetic fields might also affect  the star formation efficiency by changing the Jeans instability criteria because  the critical mass for collapse depends on the Alfven and the sound speeds.
Even if  magnetic fields play an important role in the collapse and fragmentation of cold clouds,  the generalization of Eqn. \eqref{eq9} to include magnetic fields is not a trivial task. Moreover, if we consider that the interstellar medium in dwarf galaxies is heterogeneous and has varying density, we expect hydrostatic turbulent pressure to act in differently way in different regions of a dwarf galaxy \citep{2002AJ....123.1316B}. Finally, tightly related to the studies of  star formation and  environmental effects is the problem of gas removal from  dwarf galaxies. If ram pressure and/or  gas consumption cannot explain all the present-day observations, a plausible alternative  could be the  gas ionization by UV radiation (of stellar or Galactic origin) and by SN explosions. These effects played an important role during  the epoch of reionization by increasing the the Jeans mass. Unfortunately the theoretical predictions for the low mass regime of the dwarf galaxies is strongly biased by the numerical resolution. Finally, in a cosmological context the reionization at low redshift can have played an important role in the star formation processes \citep{2011MNRAS.413.2093I}, which however we have neglected here to first approximation of the problem.

Considering our finding, a natural follow-on project would be a study of an extended family of orbits and model parameters to investigated the competitive role of the different dissipative phenomena acting around the MW, M31 or by considering the LG tidal interactions: which processes affect the star formation history, and which do not? When is a deep potential well of the dwarf galaxy sufficiently massive to suppress the effects of the various dissipative phenomena? When do tides promote or inhibit star formation through compression/stripping? Similarly for ram pressure, Rayleigh-Taylor, Kelvin-Helmholtz effects, when and where are they dominant in the history of MW dwarf galaxies?
These and many other questions can be investigated with this technique, for the first time in relation with synthetically generated colour magnitude diagrams, in order to understand how to interpret the observed CMDs (Pasetto et al. 2012, in preparation).

\begin{acknowledgements}
We thank the anonymous referee for helpful
comments. SP thanks D. Kawata for stimulating discussions and careful reading of the manuscript.
 The simulations presented in this work have been performed at the Leibnitz Rechenzentrum of Munich on the National Supercomputer HLRB II. EKG was partially supported by Sonderforschungsbereich SFB 881 "The
Milky Way System" (subproject A2) of the German Research Foundation (DFG).
 SP thanks the High Performance Computing team of the for the HLRB-II for the support in the high resolution simulation of Appendix \ref{scalability}.
\end{acknowledgements}

\appendix
\section{Testing the code}
We group in this section a few technical tests performed to control the code stability, its convergence and to compare with other literature results.

\subsection{Convergence tests and code scalability}\label{scalability}
\begin{figure}
\resizebox{\hsize}{!}{\includegraphics{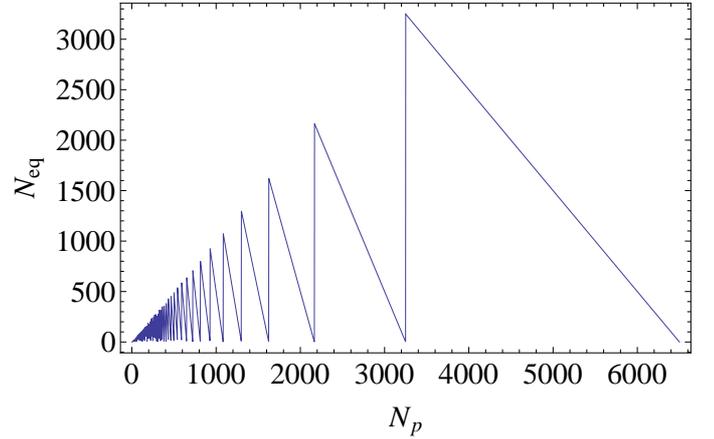}}
\caption{Plot of theoretical load balance between number of processors $N_p$ and number of equations in the system \eqref{eq1}.
}
\label{Lb}
\end{figure}

We present here the test on the convergence of the results of the integration of the system of equations proposed in Eqns. \eqref{eq1} and a few comments on a remarkable result we have obtained with the parallelization of the code.
As explained in the main text, our problem is to single out the volume occupied by the orbits (differentiable functions ${\bm{x}}:\mathbb{R} \to \mathbb{R}^3 $ of the real line (i.e., the time) on the configuration space $\mathbb{Q}$) that minimizes a $\chi ^2 $ function-difference between an observed and a synthetic CMD (convolved with distance modulus and errors). This function is clearly surjective (i.e. 'onto') so that, once the volume of the parameter space is huge and multidimensional, the possibility of rapidly converging to a best-fit solution becomes an important issue. The structure of the code we have developed is extremely promising thanks to the scalability of the system of Eqn. \eqref{eq1}. For its integration the classical Runge-Kutta method with adaptive step-size has been adopted \citep[e.g.,][]{1993Obs...113R.214P}, but the mass classes have been spread over a large number of processors. Clearly to optimize/balance the load we need to split the equations over the rank size by minimizing the latency over the available processors: $N_c  - Int\left[ {\frac{{N_c }}
{{N_p }}} \right]N_p $ with $N_p $ the number of processors and $N_c $ the number of classes. The plot is as in the Fig. \ref{Lb}.

\begin{figure}
\resizebox{\hsize}{!}{\includegraphics{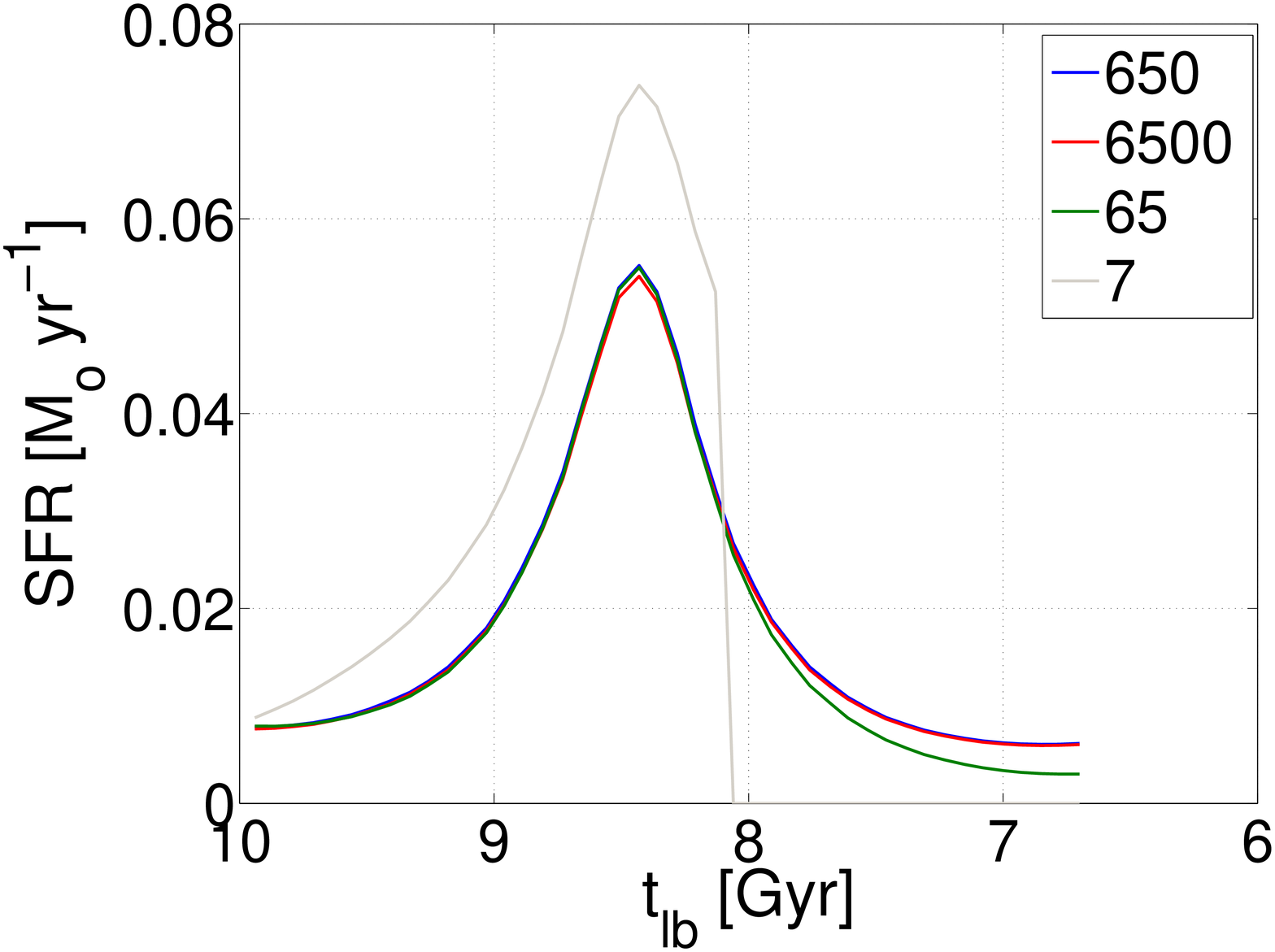}}
\caption{Plot of theoretical load balance between number of processors $N_p$ and number of equations in the system \eqref{eq1}.
}
\label{Converg}
\end{figure}

Once the performance of the code was understood, we performed a few convergence tests on the astrophysical results. Depending on the available computational resources, say the $N_p$, we can choose the best number of processors on which to perform our integration by looking at Fig. \ref{Lb}. We had the possibility to test the lower minimum peak on the right of Fig. \ref{Lb} (for $N_{p}=6500$). For this high resolution test, the system of equations of Eqn. \eqref{eq1} has been integrated with $\alpha  = \frac{1}{{1000}}$ as defined in Section \ref{Molecular clouds evolution and non gravitational heating}.
The results are presented in Fig. \ref{Converg} where four lines are plotted, for the integration of the system of equation \eqref{eq1} with 7, 65, 650, 6500 molecular cloud classes on an E050 orbit. The test shows how the approximation adopted in our study, for $\alpha  = 1/100$, i.e., 650 molecular classes (the blue line), gives a robust consistent star formation rate with the higher resolution simulation (the red line): red and blue line overlap almost perfectly.

On 1st May 2011 the simulation ran on 6500 processors at SGI Altix 4700 at the National Supercomputer HLRB-II, Munich ( Germany) on a dedicated queue recording a peak performance of 23Tflops/second.

\subsection{Tests on the integro-differential system of equations Eqn. \eqref{eq1}}\label{systen9}\label{SFHsystem}
\begin{figure}
\resizebox{\hsize}{!}{\includegraphics{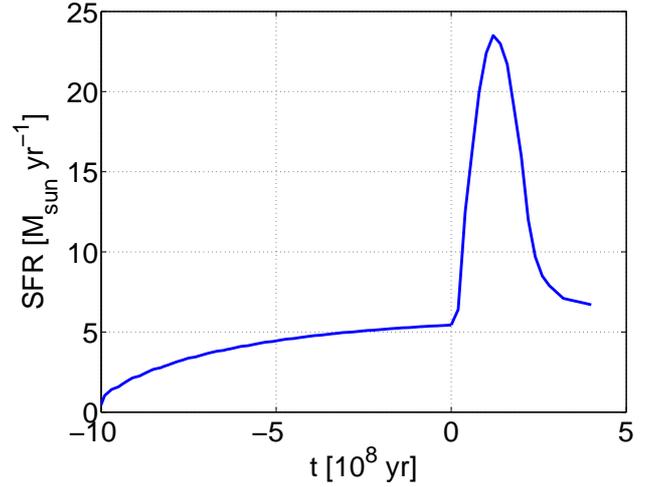}}
\caption{Star formation rate produced by the pressure injected from Eqn. \ref{PressureFujita}}.
\label{Fujitatest}
\end{figure}

Eqn. \eqref{eq1} is an integro-differential system of equations used to deduce the star formation rate of a galaxy from the total pressure $P$. While this system is conceptually stable because it is based on the fundamental principles of stellar structure and evolution \citep[e.g.,][]{2008A&A...484..815B,2009A&A...508..355B}, its integration is not trivial and the numerical approaches adopted needs to be tested.
Its integration on the independent variable $t$ is the most computational time consuming part of the code and has been performed with distributed-memory parallelization standard techniques (message-passing interface, MPI). A serial integration of a similar system has been already performed in literature by \citet{1998ApJ...509..587F}.
Therefore, we can prepare a case-test to reproduce the same result posted in \citet{1998ApJ...509..587F} by omitting the pressure determination with our Eqn. \eqref{eq9} and instead by ''injecting'' in our code the synthetic pressure profile of Fig.1a (dashed line) in \citet{1998ApJ...509..587F}. Thus, we eliminate the code section devoted to our original pressure determination and only the parallel integrator of the system of equation Eqn.\eqref{eq1} is tested against the serial independent determination originally presented in \citet{1998ApJ...509..587F}. The pressure is defined piece-wise as follow:
\begin{equation}\label{PressureFujita}
	P\left( t \right) = \left\{ {\begin{array}{*{20}c}
	   {P_ \odot  } & {t \le 0}  \\
	   {P_ \odot  e^{\frac{3}{2}\left( {\ln 2 + \ln 5} \right)t} } & {0 < t \le 2}  \\
	   {1000P_ \odot  } & {t > 2.}  \\
	\end{array}} \right.
\end{equation}
We see in Fig.\ref{Fujitatest} that our results that are almost identical to the dashed line of Figure 1b in \citet{1998ApJ...509..587F}. Here the integration is performed with Runge-Kutta method with adaptive stepsize control \citep[e.g.,][]{1993Obs...113R.214P} and distributed over 650 processors.  

\subsection{Pressure from Eqn.\eqref{eq9} and and speed limits}\label{pressureprofiles}
\begin{figure}
\resizebox{\hsize}{!}{\includegraphics{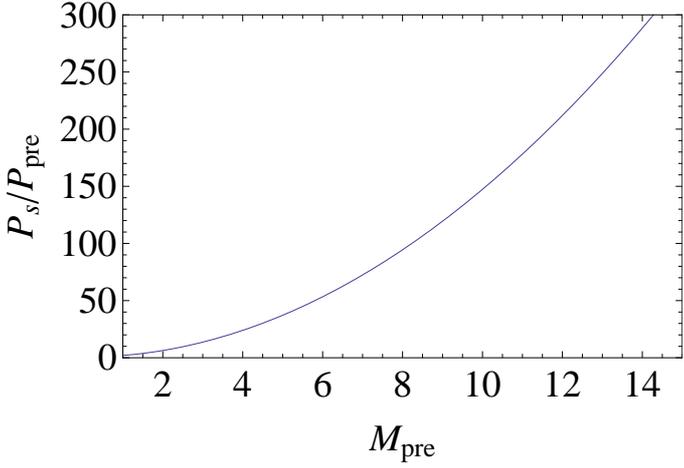}}
\caption{Tend between of the ratio between pressure at the stagnation point and pressure pre shock as a function of the Mack number of the gas where the dwarf galaxy is moving.}
\label{superMach}
\end{figure}

We assume here that the dwarf galaxy is falling in an general intra cluster medium (ICM) or that a dwarf galaxy is passing through the MW gaseous disk. In these situations the motion of the dwarf can easily be in supersonic regime, i.e. Mach number ${\rm M}_{{\text{pre}}}  > 1$, or equivalently that the flow is impacting on our dwarf galaxy at supersonic velocity $v_{\text{g}}  > v_{s,{\text{ICM}}} $
 with $v_{s,{\text{ICM}}} $ sound speed of the flow. Moreover, we keep general this exercise assuming a polytrophic equation of state equation with adiabatic index $\gamma $
 for the gas flowing. The presence of a shock increase density (and pressure) by compression while the velocity field from supersonic become subsonic (we will limit our arguments to normal shocks) whence Eqn.\eqref{eq9} holds. Nevertheless, it's simple to prove that the pressure of the gas before the bow shock, $P_{{\text{pre}}} $, and the stagnation point  $P_{s} $ can be very different and the Eqn.\eqref{eq9} have to be 'clothed' with supersonic formalism. We call
 the pressure after the surface of discontinuity (the thin shock), $P_{{\text{post}}} $. In this case the relations between pressures pre and post shock are already known \cite[e.g.,][]{1959flme.book.....L}
\begin{equation}\label{shockseq}
	\begin{gathered}
	  \frac{{P_{{\text{post}}} }}
	{{P_{{\text{pre}}} }} = 1 + \frac{{2\gamma }}
	{{\gamma  + 1}}\left( {{\rm M}_{{\text{pre}}}^2  - 1} \right) \hfill \\
	  {\rm M}_{{\text{post}}}^2  = \frac{{\gamma  + 1 + \left( {\gamma  - 1} \right)\left( {{\rm M}_{{\text{pre}}}^2  - 1} \right)}}
	{{\gamma  + 1 + 2\gamma \left( {{\rm M}_{{\text{pre}}}^2  - 1} \right)}} \hfill \\
	\end{gathered}
\end{equation}
and we obtain the pressure between $P_{{\text{post}}} $
 and $P_s $ as
\begin{equation}
	\frac{{P_s }}
	{{P_{{\text{post}}} }} = \left( {\frac{{\gamma  - 1}}
	{2}{\rm M}_{{\text{post}}}^2  + 1} \right)^{\frac{\gamma }
	{{\gamma  - 1}}} \end{equation}
With ${\rm M}_{{\text{post}}}^2 $  and $\frac{{P_{{\text{post}}} }}{{P_{{\text{pre}}} }}$
 from Eqn.\ref{shockseq} we can write the relation between $\frac{{P_s }}{{P_{{\text{pre}}} }}$
 as:
\begin{equation}
\begin{gathered}
  \frac{{P_s }}
{{P_{{\text{pre}}} }} = \frac{{P_s }}
{{P_{{\text{post}}} }}\frac{{P_{{\text{post}}} }}
{{P_{{\text{pre}}} }} \\
   = \left( {\frac{{\gamma  + 1}}
{2}{\rm M}_{{\text{pre}}}^2 } \right)^{\frac{\gamma }
{{\gamma  - 1}}} \left( {1 + \frac{{2\gamma }}
{{\gamma  + 1}}\left( {{\rm M}_{{\text{pre}}}^2  - 1} \right)} \right)^{ - \frac{1}
{{\gamma  - 1}}}  \\
\end{gathered}
\end{equation}
whose plot is as in Fig.\ref{superMach}.
As evident, the pre-shock pressure and the pressure at the stagnation point are not of the same order even if the post shock velocity field is subsonic: the flow is compressed and its pressure (and density) increases passing through the shock and after the shock before reaching the stagnation point. As seen in Fig. \ref{superMach} these pressure remain of the same order only for subsonic or weakly supersonic that represent the limit of validity of Eqn.\eqref{eq9}, before supersonic correction has to be implemented.

\section{Star formation efficiency and disruption time relation}\label{SFMC}
The connection between star formation processes and molecular clouds is a very fertile research topic encompassing observations \citep[e.g.,][]{2008AJ....136.2846B}, theory \citep[e.g.,][]{2005ApJ...630..250K} and experimental/numerical works \citep[e.g.,][]{2011arXiv1109.0467K}. In our approach we revisited a work of \citet{1997ApJ...480..235E} chosen for its simplicity. Other literature results can be similarly implemented.
Let $M_i $ be the cloud's initial mass as defined in Section \ref{Molecular clouds evolution and non gravitational heating} for which we assumed a constant star formation rate $\varsigma  _i $ in the integration interval $dt$ \citep[see also][]{1989ApJ...338..178E, 2005ApJ...630..250K, 2007ApJ...654..304K}. Following \citet{1989AN....310..381S}, the erosion rate is proportional to the luminosity $L_s $ of embedded stars with total mass $M_{{\text{star}},i} $ divided by the specific mass binding energy $M_i \sigma ^2 $  with $\sigma $ dispersion velocity. The equation for the rate of change of gas mass in the cloud considered is:
\begin{equation}\label{A1}
	\frac{{dM_i }}
{{dt}} =  - \varsigma  _i  - A_i \frac{{L_i }}
{{\sigma ^2 }};
\end{equation}
the luminosity
\begin{equation}\label{A2}
	L_i \left( t \right) = \int_0^t {\varsigma  _i \Lambda _i \left( {t - t'} \right)dt'},
\end{equation}
where $\Lambda _i \left( t \right)$  is the luminosity-to-mass ratio of a population of stars generated by the cloud class at the instant $t$ that we obtained as explained in Section \ref{Synthetic colour magnitude diagrams}.
Let us drop for the moment the subscript $i$ to simplify the notation. From the previous Eqns. \eqref{A1} and \eqref{A2} we get:
\begin{equation}\label{A3}
\frac{{dM  }}
{{dt}} =  - \varsigma    \left( {1 + \frac{{A  }}
{{\sigma ^2 }}\int_0^t {\Lambda   \left( {t - t'} \right)dt'} } \right),
\end{equation}
which can be integrated to give
\begin{equation}\label{A4}
	M  \left( t \right) = M  \left( 0 \right) - \varsigma    \left( {t + \frac{{A  }}
{{\sigma ^2 }}\int_0^t {dt'\int_0^{t'} {\Lambda   \left( {t' - t''} \right)dt''} } } \right),
\end{equation}
where the double integral is more easily numerically computed as:
\begin{equation}\label{A5}
\begin{gathered}
  \int_0^t {dt'\int_0^{t'} {\Lambda   \left( {t' - t''} \right)dt''} }  =  \hfill \\
  t\int_0^t {\Lambda   \left( {t - t'} \right)dt' - \int_0^t {t'\Lambda   \left( {t'} \right)dt'} }  \hfill \\
\end{gathered}
\end{equation}
to give:
\begin{equation}\label{A6}
\begin{gathered}
  M  \left( t \right) = M  \left( 0 \right) - \varsigma  \left( {t + \frac{{A  }}
{{\sigma ^2 }}\left( {t\int_0^t {\Lambda   \left( {t - t'} \right)dt'} } \right.} \right. \hfill \\
  \left. {\left. { - \int_0^t {t'\Lambda   \left( {t'} \right)dt'} } \right)} \right) \hfill \\
\end{gathered}
\end{equation}
for every $i$. We can find also a quicker approach based on an interpolation function in the work of \citet[][their Fig. 13]{1995A&A...298...87G} as recently in \citet{2006ApJ...644..879E} that we can also adopt when necessary to speed up the code. By using a power-law like $\Lambda \left( t \right) = \Lambda _0 \left( {\frac{t}
{{t_0 }}} \right)^{ - \lambda } $,  $\lambda  \in \left[ {0,1} \right[$, the integrals in Eqn. \eqref{A6} can be carried out analytically. We observe that the destruction time $\tau $
of the molecular clouds can be defined in an implicit way as the instant $\tilde t|M  \left( {\tilde t} \right) = 0$:
\begin{equation}\label{A7}
	\begin{gathered}
   - \tilde t\varsigma     - \frac{{A  \tilde t^2 \Lambda _0 \varsigma    }}
{{\left( {\lambda  - 1} \right)\left( {\lambda  - 2} \right)\sigma ^2 }}\left( {\frac{{\tilde t}}
{{t_0 }}} \right)^{ - \lambda }  + M  \left( 0 \right) = 0 \Leftrightarrow  \hfill \\
  1 - \alpha \tau  - \beta \frac{{\alpha ^2 }}
{{\left( {\lambda  - 1} \right)\left( {\lambda  - 2} \right)}}\tau ^{2 - \lambda }  = 0, \hfill \\
\end{gathered}
\end{equation}
where we exploited the following definitions:
\begin{equation}\label{A8}
	\tau  \equiv \frac{t}
{{t_0 }},
\end{equation}
and
\begin{equation}\label{A9}
	\begin{gathered}
  \alpha  \equiv \frac{{t_0 \varsigma    }}
{{M  \left( 0 \right)}} \hfill \\
  \beta  \equiv \frac{{A  \Lambda _0 }}
{{\sigma ^2 }}\frac{{M  \left( 0 \right)}}
{{\varsigma    }}, \hfill \\
\end{gathered}
\end{equation}
with $t_0  \sim 10^7 $ yr as a fixed parameter from \citet{1995A&A...298...87G} and $\lambda  \sim 0.6$
 for $t > t_0 $
 and $\lambda  = 0$
 for $t < t_0 $
where the solution is analytical (a quadratic equation).  Apart from the difference in the mathematical formalism here laid out, the contents follow then exactly as in \citet{1997ApJ...480..235E} to which we refer the reader for extended discussion. We point out here that we can define the efficiency as
\begin{equation}\label{A10}
	\varepsilon    \equiv \frac{{\varsigma    }}
{{M  \left( 0 \right)}}t.
\end{equation}
The two functions defined in \eqref{A8} are of interest due to their dependence on pressure impacting on the orbital mass of the dwarf galaxies in our LG. We assume the following functional dependence \citep{1989ApJ...338..178E, 1997ApJ...480..235E}:
\begin{equation}\label{A11}
	\begin{gathered}
  \alpha  = \alpha \left( {M,P} \right) = \alpha _0 \left( {\frac{M}
{{10^5 M_ \odot  }}} \right)^{ - 1/4} \left( {\frac{P}
{{P_ \odot  }}} \right)^{3/8}  \hfill \\
  \beta  = \beta \left( {M,P} \right) = \beta _0 \left( {\frac{M}
{{10^5 M_ \odot  }}} \right)^{ - 1/4} \left( {\frac{P}
{{P_ \odot  }}} \right)^{ - 5/8},  \hfill \\
\end{gathered}
\end{equation}
with dimensionless constants $\alpha _0  = 0.1$, $\beta _0  = 180$, $P_ \odot   = 3 \times 10^4 k_B {\text{cm}}^{ - 3} {\text{K}}$, and $k_B$ as Boltzmann constant.

% for the bibliography, at the end
\bibliographystyle{aa} % style aa.bst
\bibliography{BiblioArt} % your references Yourfile.bib
\end{document}